\documentclass[twocolumn,secnumarabic,amssymb, nobibnotes, aps, prx,superscriptaddress]{revtex4-1}

\usepackage[paperwidth=210mm,paperheight=297mm,centering,hmargin=2cm,vmargin=2.5cm]{geometry}
\usepackage{amsmath} 
\usepackage[english]{babel} 
\usepackage{graphicx} 
\usepackage{listings} 
\usepackage{soul}
\usepackage{color}
\usepackage{lipsum}
\usepackage{float}

\begin{document}

\title{Microgels at interfaces behave as 2D elastic particles featuring reentrant dynamics}

\author{Fabrizio Camerin}
\email[Corresponding author:]{ fabrizio.camerin@uniroma1.it}
\affiliation{CNR Institute of Complex Systems, Uos Sapienza, Piazzale Aldo Moro 2, 00185 Roma, Italy}
\affiliation{Department of Basic and Applied Sciences for Engineering, Sapienza University of Rome, via Antonio Scarpa 14, 00161 Roma, Italy}

\author{Nicoletta Gnan}
\affiliation{CNR Institute of Complex Systems, Uos Sapienza, Piazzale Aldo Moro 2, 00185 Roma, Italy}
\affiliation{Department of Physics, Sapienza University of Rome, Piazzale Aldo Moro 2, 00185 Roma, Italy}

\author{Jos\'e Ruiz-Franco}
\affiliation{Department of Physics, Sapienza University of Rome, Piazzale Aldo Moro 2, 00185 Roma, Italy}
\affiliation{CNR Institute of Complex Systems, Uos Sapienza, Piazzale Aldo Moro 2, 00185 Roma, Italy}

\author{Andrea Ninarello}
\affiliation{CNR Institute of Complex Systems, Uos Sapienza, Piazzale Aldo Moro 2, 00185 Roma, Italy}
\affiliation{Department of Physics, Sapienza University of Rome, Piazzale Aldo Moro 2, 00185 Roma, Italy}

\author{Lorenzo Rovigatti}
\affiliation{Department of Physics, Sapienza University of Rome, Piazzale Aldo Moro 2, 00185 Roma, Italy}
\affiliation{CNR Institute of Complex Systems, Uos Sapienza, Piazzale Aldo Moro 2, 00185 Roma, Italy}

\author{Emanuela Zaccarelli}
\email[Corresponding author:]{ emanuela.zaccarelli@cnr.it}
\affiliation{CNR Institute of Complex Systems, Uos Sapienza, Piazzale Aldo Moro 2, 00185 Roma, Italy}
\affiliation{Department of Physics, Sapienza University of Rome, Piazzale Aldo Moro 2, 00185 Roma, Italy}

\date{\today}

\begin{abstract}
\noindent

Soft colloids are increasingly used as model systems to address fundamental issues such as crystallisation and the glass and jamming transitions. Among the available classes of soft colloids, microgels are emerging as the gold standard. Since their great internal complexity makes their theoretical characterization very hard, microgels are commonly modelled, at least in the small-deformation regime,
within the simple framework of linear elasticity theory.
Here we show that there exist conditions where its range of validity
can be greatly extended, providing strong numerical evidence that microgels adsorbed at an interface follow the two-dimensional Hertzian theory, and hence behave like 2D elastic particles, up to very large deformations in stark contrast to what found in bulk conditions. 
We are also able to estimate the Young's modulus of the individual particles and, by comparing it with its counterpart in bulk conditions, we demonstrate a significant stiffening of the polymer network at the interface.
Finally, by analyzing dynamical properties, we predict multiple reentrant phenomena: by a continuous increase of particle density, microgels first arrest and then re-fluidify due to the high penetrability of their extended coronas. We observe this anomalous behavior in a range of experimentally accessible conditions for small and loosely crosslinked microgels.
The present work thus establishes microgels at interfaces as a new model system for fundamental investigations, paving the way for the experimental synthesis and research on unique high-density liquid-like states. In addition, these results can guide 
the development of novel assembly and patterning strategies on surfaces and the design of novel materials with desired interfacial behavior.
\end{abstract}

\maketitle

\section{Introduction}
Mesoscopic assemblies of colloids and nanoparticles display features that depend critically on the microscopic details of the building blocks, \textit{e.g.} composition, size and shape, as well as on the specific macroscopic physical conditions such as the thermodynamic control parameters. It is by carefully choosing and tuning these variables that one can induce the formation of different structures and explore various states, such as liquid-like fluids, glasses or crystals~\cite{lubchenko2007theory}. At the core of this collective behaviour is the interparticle interaction, which ultimately dictates the phase behaviour and dynamics of the assembly: information at the multi-particle level is thus fundamental to determine the properties of the material.
If a system is made of rigid building blocks that interact only through excluded volume interactions, it can be approximately mapped to a hard-sphere system and its behaviour can be investigated through packing models. These have been used for a long time to successfully answer fundamental questions in physics and material science whenever simple constituent units are involved~\cite{torquato2018perspective}. 
\\
\indent
However, in certain cases, the complexity that resides at the microscopic level cannot be described in these terms. This is especially true for soft polymeric colloids that possess internal degrees of freedom endowing them with elasticity and deformability. Among the available library of soft deformable particles, microgels, colloidal-sized crosslinked polymer network, are one of the finest illustrations of this concept
~\cite{lyon2012polymer,karg2019nanogels,martin2019review,oberdisse2020recent,rey2020poly}. Their 
structure is determined by the chemical synthesis conditions that, in the common
procedure of precipitation polymerization~\cite{pelton2000temperature}, lead to the formation of spherical particles made of a compact core and a fluffy external corona~\cite{siemes2018nanoscopic}. Although microgels are often considered as simple elastic particles that can be modeled with the classical elasticity theory, the presence of multiple length scales in their internal architecture makes their effective interactions in bulk more complex than a pure Hertzian model and calls for more refined treatments that range from a phenomenological multi-Hertzian model~\cite{bergman2018new} to descriptions that depend on the concentration regime~\cite{scheffold2010brushlike,conley2019relationship}. 
\\
\indent
The intrinsic softness of microgels and, in general, of soft deformable objects, is fully revealed, and can be taken advantage of, at interfaces, which can be used to fulfil different  purposes. In fact, if the interfacial tension is large enough,
it is possible to coat an interface with nano- or microsized particles which then remains adsorbed and can form stable monolayers. This concept can be used, for instance, to stabilize emulsions and biological membranes~\cite{monteillet2014ultrastrong,kwok2019microgel,wang2019assembling,tatry2019kinetics,richtering2012responsive} or to study fundamental self-assembly phenomena in 2D such as crystal~\cite{rey2016isostructural} or quasi-crystal formation~\cite{dotera2014mosaic,zu2017forming}. 
Moreover, by changing \textit{in situ} the single-particle properties and the local environment, it is possible, in principle, to finely control the stability and the structure of the whole monolayer/emulsion~\cite{schmidt2011influence,liu2012non,grillo2019self,volk2019moire}. Noteworthy, in this respect, is the possibility to realize complex patterns whose application as etching masks in nanolithography can lead, for instance, to the fabrication of nanowire arrays~\cite{fernandez2018tunable,rey2015fully}.
\begin{figure}[t!]
\centering
\includegraphics[scale=0.3]{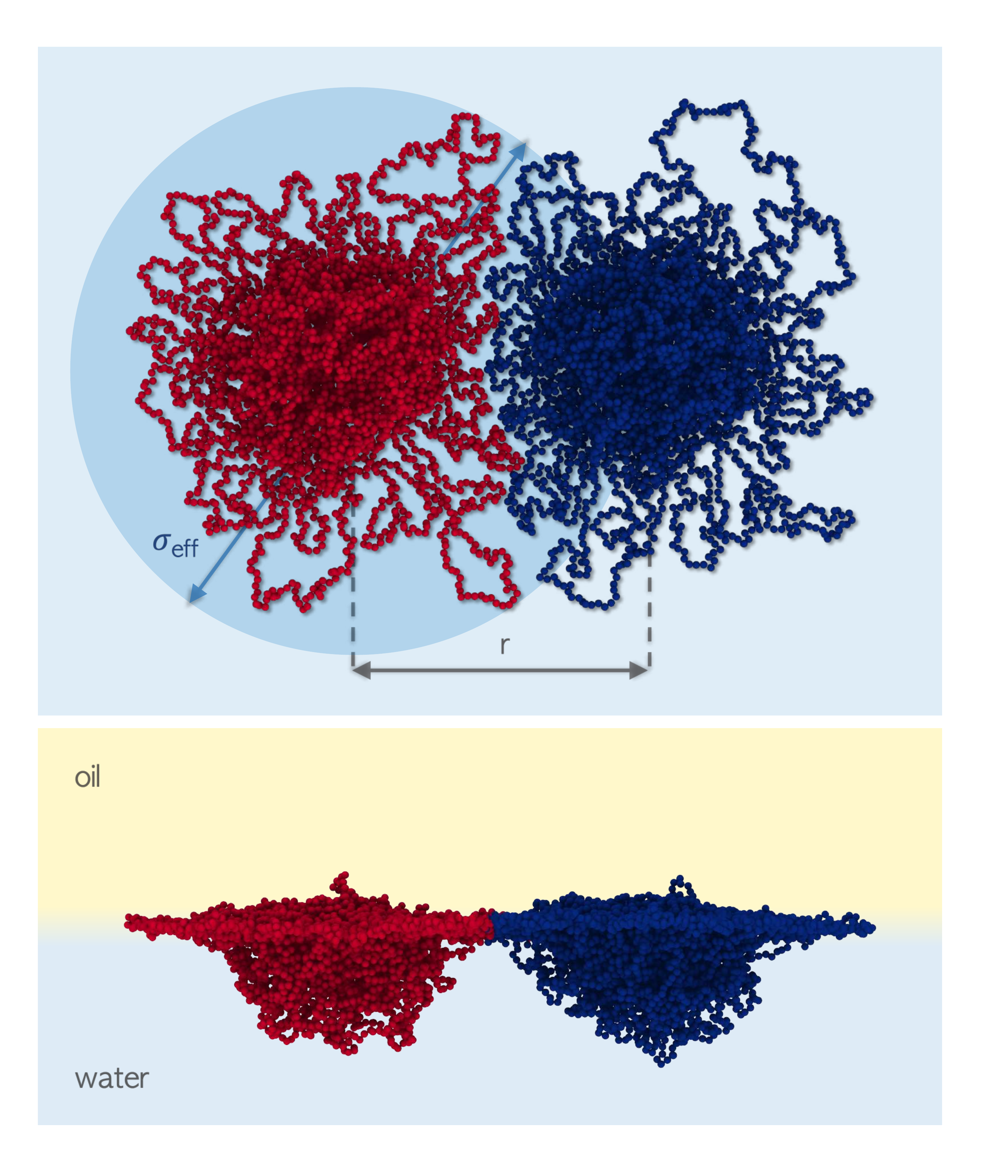}
\caption{\textbf{Microgels interacting at the interface.} Top and side simulation snapshots of two microgels with $c=5\%$ at the water-oil interface at a representative distance $r\approx 40 \sigma_{\rm m}$. The effective diameter of the microgel is $\sigma_{\rm eff}$. Solvent particles are not shown for clarity.
}
\label{fig:fig1_snaps}
\end{figure}
\\
\indent
When adsorbed at interfaces, microgels flatten out and adopt the so-called ``fried-egg'' shape, making them very different from their bulk counterparts~\cite{geisel2012unraveling,scotti2019exploring,camerin2019microgels}. In particular, at an interface, the polymer network tends to minimize as much as possible the surface tension between the two liquids by taking a stretched configuration already in the dilute regime.
So far, the effects that an interfacial confinement induces on such particles 
have been mainly limited to the study of their characteristic shape, and hence to their structural arrangement at oil-water and air-water interfaces. These aspects have been widely investigated experimentally~\cite{maldonado2017thermoresponsive,agrawal2018stimuli} and, more recently, also numerically~\cite{camerin2019microgels,harrer2019stimuli,arismendi2020deformability}. 
\\
\indent
By contrast, little is known on the collective behavior of such particles,
mostly by means of indirect experimental feedback~\cite{deshmukh2014equation,bochenek2019effect,picard2017organization,scheidegger2017compression,rey2017anisotropic}, which has neither allowed to extract a functional form for the interaction potential nor to properly understand the role of the surface tension. In the same way, a true characterization of the elasticity of the polymer network within the interfacial plane is still missing, being limited both experimentally and numerically by subpar techniques and models. Clearly, a simple transfer of results from bulk to interface would be highly inappropriate, due to the dramatic change of conditions between the two cases.
\\
\indent
This lack of understanding hampers the progress towards further applications, since an established fundamental knowledge of the basic constituents would make it possible to \textit{a priori} design and guide the assembly of innovative materials and nanostructures. From a theoretical standpoint, it also prevents the adoption of microgels at interfaces as model systems for the study of open questions in fundamental science~\cite{maestro2019tailoring,deshmukh2015hard}. In this sense, it is crucial to provide a microscopic understanding of such system.
\\
\indent
In this work, we address this problem by calculating both the effective interactions between two microgels at a liquid-liquid interface and their individual elastic properties, using this knowledge to predict their multi-particle response at high densities.
Our approach relies on state-of-the-art modeling of single-particle microgels that was shown to quantitatively capture the internal topology of laboratory microgels both in the bulk~\cite{ninarello2019modeling} and at the interface~\cite{camerin2019microgels}. In the latter case an appropriate framework, that explicitly takes into account the effects of the surface tension between the two liquids, has been developed in order to correctly describe the deformation of the microgels~\cite{camerin2019microgels}.
\\
\indent
Despite the complex arrangement of the polymer network at the interface and the intrinsic presence of a core-corona structure, the calculation of the effective interactions between two microgels 
on the interfacial plane reveals a remarkable agreement with the 2D expression of the Hertzian model for elastic disks for all investigated distances and crosslinker concentrations $c$. This is clearly different from what was found for the same system in bulk conditions~\cite{rovigatti2019connecting} and establishes the validity of the two-dimensional Hertzian model for microgels at interfaces up to large compression regimes. The Young's modulus determined from the effective potential is also directly compared to explicit calculations 
based on elasticity theory for small and intermediate deformations. 
Thanks to this method we are able to achieve a full characterization of the elastic response of the microgels in the two-dimensional interfacial plane and we can thus establish a sound comparison to the three-dimensional bulk case. 
Notably, our results show that the elastic moduli, once converted to their three-dimensional counterparts, are roughly one order of magnitude larger at the interface than for the same microgels in the bulk.
This highlights the key role of the interfacial tension in stiffening the microgels due to the stretching of their coronas. 
\\
\indent
Having determined how such complex particles interact with each other, we are finally able to carry out our study also at the collective level by investigating the dynamical phase behavior of an ensemble
of these effective elastic disks. When softness and elasticity are taken into account in the interparticle interaction a rich behavior is, in general, expected~\cite{vlassopoulos2014tunable,lacour2019influence,gnan2019microscopic}. In particular, we find the presence of multiple reentrant melting phenomena, where a glass is melted simply by an increase in particle concentration. Although similar findings have long been predicted for simple soft models~\cite{berthier2010increasing}, here, for the first time, such scenario is found for microscopically-motivated effective interactions and, most importantly, for potential parameters that can be realized in experiments.
\\
\indent
The extensive analysis of microgels at the interface presented here, together with the notion that the Hertzian model can be used up to very large deformation energies, sheds light on how single microgel properties and collective response are coupled, and demonstrates that this system is an extremely promising model for probing the collective behaviour of 2D elastic particles up to large densities. Their unusual dynamical features are of wide interest for the preparation of colloidal monolayers with non-monotonic viscoelastic properties that could be used for a variety of different applications.

\section{The model}
Microgel monomers interact through the Kremer-Grest bead-spring model~\cite{kremer1990dynamics,grest1986molecular} in explicit solvent. The polymer network has a disordered topology~\cite{gnan2017silico} and the distribution of crosslinker is inhomogeneous, slowly rarefying towards the corona, as typically results from microgels synthesized via precipitation polymerization. These factors allow for a favorable comparison to realistic microgels, both in terms of form factors and density profiles~\cite{ninarello2019modeling}.
The solvent is treated within the Dissipative Particle Dynamics framework~\cite{camerin2018modelling} and, for simulations at the interface, their mutual interactions are tuned to mimic a water/hexane interface. The surface tension between the two fluids is representative of a wide variety of solvents that are typically employed for such studies at the interface. Furthermore, no relevant differences in the distribution of the microgel monomers on the plane of the interface is expected by changing the value of the surface tension~\cite{camerin2019microgels}.
\\
\indent
Under these conditions the microgel  spontaneously adopts the characteristic ``fried-egg" shape when placed close to the interfacial plane. Its structural characterization is in good agreement with experiments both in terms of flattening on the interfacial plane and of protrusion on the preferred water side~\cite{camerin2019microgels}. In the present work, for the study of the microgel-microgel effective interactions, we simulate particles with $N\approx 5000$ monomers of diameter $\sigma_{\rm m}$, that defines the unit of length, and crosslinker molar fraction $c=3\%, 5\%$ and $10\%$.
 Due to the exceptional computational cost to carry out the simulations, we limit our study to a single microgel topology for each studied value of $c$, checking the consistency of the results with a second topology for $c=5\%$ (see Supplemental Material). The elastic properties of smaller microgels with 2000 and 3000 monomers are also analyzed for further considerations on their collective behavior.
Additional details on the microgel and interface modeling, and on simulations are provided in the Methods section and in Refs.~\cite{camerin2019microgels,camerin2018modelling,ninarello2019modeling}. The typical conformation taken by two interacting microgels at the interface is reported in the simulation snapshots 
of Fig.~\ref{fig:fig1_snaps}. 

\section{Results and Discussion}
\begin{figure}[!t]
\centering
\includegraphics[scale=0.45]{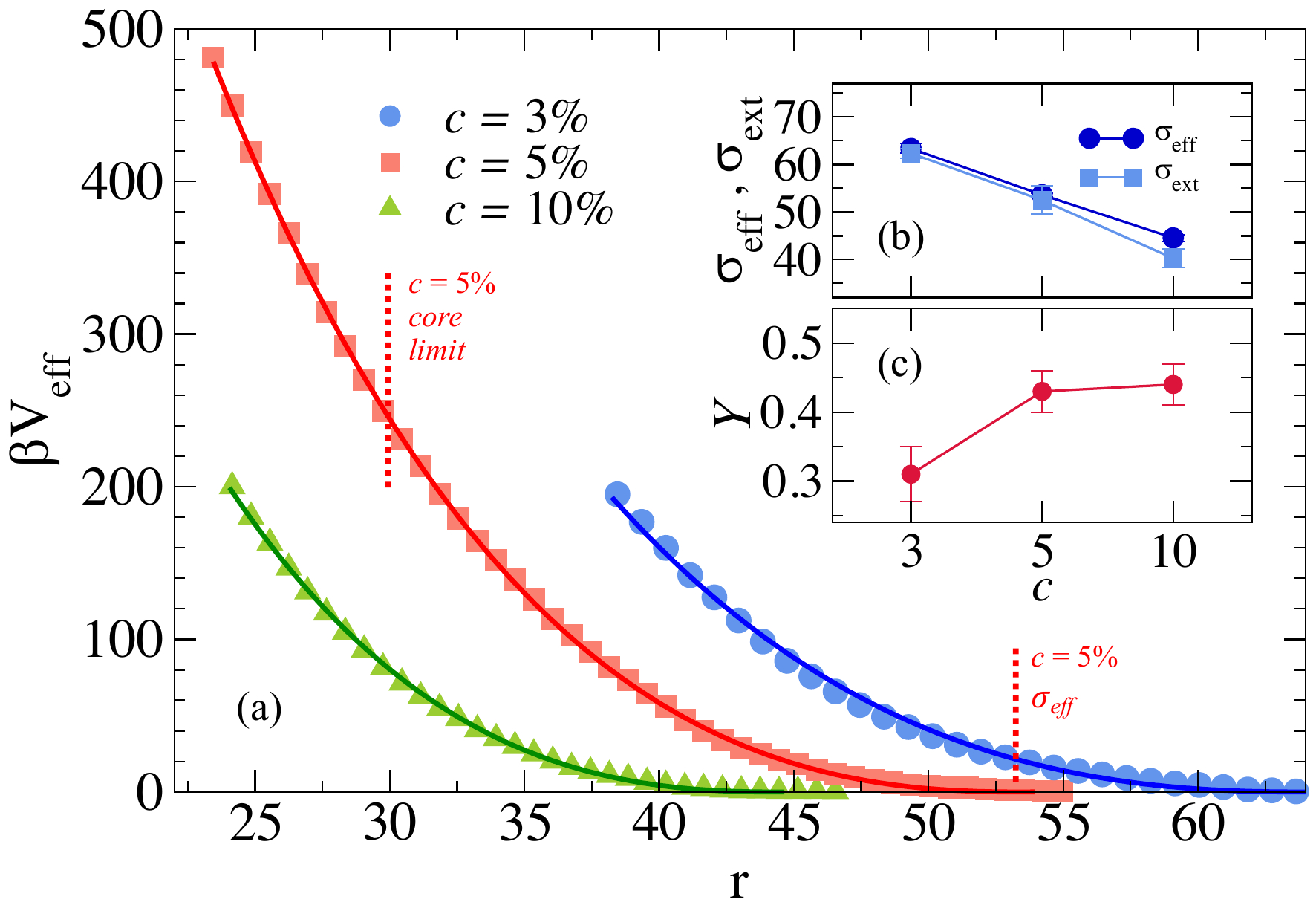}
\caption{\textbf{Effective potentials for microgels at the liquid-liquid interface and related 2D Hertzian fit parameters.} (a) Numerical results refer to three values of the crosslinker concentrations $c$: $3\%$ (circles), $5\%$ (squares) and $10\%$ (triangles). Full lines are fits to numerical results using Eq.~\ref{2dhertz}. 
The two vertical dotted lines indicate the value $\sigma_{\rm eff} \approx 53\sigma_{\rm m}$ and the distance at which the cores of the two microgels get in contact ($r\approx30\sigma_{\rm m}$) for $c=5\%$; (b) microgel effective diameter $\sigma_{\rm eff}$ compared to the extension on the plane of the interface $\sigma_{\rm ext}$ calculated as in Ref.~\cite{camerin2019microgels}, in units of $\sigma_{\rm m}$; (c) Young's modulus $Y$ extracted from the 2D Hertzian fit to $V_{\rm H}$ in units of $k_BT/\sigma_{\rm m}^2$.}
\label{fig:fig2_effpot_interf}
\end{figure}
\subsection{Effective interaction potential}

The two-body effective potential $V_{\rm eff}(r)$ between the microgels at a water/hexane interface is evaluated by means of extensive simulations exploiting the Umbrella Sampling technique~\cite{blaakaccurate,likos2001effective,gnan2014casimir}, as also explained in the Methods section, and it is shown in Fig.~\ref{fig:fig2_effpot_interf}(a) for all investigated values of $c$, rescaled by $\beta = 1 / k_BT$.
The numerical results are compared to the two-dimensional Hertzian expression~\cite{gerl1999coefficient,hayakawa2002simulation} that reads as
\begin{equation}\label{2dhertz}
V_{\rm H}(r)=\frac{\frac{1}{2}\pi Y \sigma_{\rm eff}^2 \left(1-\frac{r}{\sigma_{\rm eff}}\right)^2}{\ln\left(\frac{2}{1-\frac{r}{\sigma_{\rm eff}}}\right)}
\end{equation}
where $r$ is the distance between the centers of mass of the microgels at the interface, $\sigma_{\rm eff}$ quantifies the effective microgel diameter on the interfacial plane and $Y$ is the Young's modulus of the individual particle.
The agreement between the numerical results and the theoretical fits is remarkable for all probed distances and all studied values of $c$. Therefore, it clearly emerges from these findings that two microgels confined at an interface effectively behave as 2D elastic objects, further confirming the soft repulsive nature of their mutual interactions. Further discussion on the functional form of the potential can be found in the Supplemental Material.
\\
\indent
Experimentally, small microgels -- having a diameter $\lesssim 200$ nm
-- are the best candidates to interact in this way, since they do not experience long-range attractions due to capillary effects~\cite{scheidegger2017compression}. Indeed, the latter have been widely reported~\cite{rey2017interfacial,huang2017structure,el2018multiple} and found to be relevant only for microgels large enough to induce a local deformation of the water/oil interface~\cite{scheidegger2017compression}. By contrast, our solvent modeling is aimed essentially at reproducing the surface tension and the microgels solubility, both of which have a direct influence on the conformation of the particle. We can thus directly probe the elastic interactions between the microgels without the interference of attractive capillary forces.
\\
\indent
These outcomes also evidence the presence of a single characteristic length in the potential up to a center-to-center distance as small as the interaction radius of the microgel ($\sim\sigma_{\rm eff}/2$) for the case $c=5\%$, which we have probed up to a repulsion of $\approx 500 k_BT$.  
The observed behavior is strikingly different from the corresponding bulk one, where the Hertzian potential was found to be valid only up to a few $k_B T$s~\cite{rovigatti2019connecting}. Indeed, in bulk, the distinction between core and corona imposes to consider different kind of interactions, depending on the investigated distances~\cite{bergman2018new}, that would describe different inner regions of the particle with changing elastic properties. Instead, at an interface, the microgel behaves as if the polymer network were more homogeneous and uniform, as indicated by the continuous and steady growth of the potential that persists even inside the core region, here corresponding to $r\lesssim 30\sigma_{\rm m}$ for $c=5\%$, as also reported in Fig.~\ref{fig:fig2_effpot_interf}(a). This behavior suggests a dominant role of the surface tension which completely controls the properties of the microgel at the interface, so that even the part of the core that protrudes from the plane of the interface effectively contributes to the 2D Hertzian description.
Thus, microgels adsorbed at interfaces represent the first colloidal system to behave as an ideal Hertzian model, when considered as two-dimensional objects on the interfacial plane. Their behavior is thus clearly different from that of microgels in bulk~\cite{bergman2018new,rovigatti2019connecting}.
\\
\indent
By fitting the calculated potential with Eq.~\ref{2dhertz}, we can obtain the effective diameter $\sigma_{\rm eff}$ of the flattened microgel and  its Young's modulus $Y$.  Interestingly, the latter quantity can be also directly estimated from the fit of the calculated potentials, at odds with the corresponding 3D case where two non-independent elastic parameters, namely $Y$ and the Poisson's ratio $\nu$, are contained in the Hertzian prefactor~\cite{rovigatti2019connecting}. The resulting fit parameters are shown in Fig.~\ref{fig:fig2_effpot_interf}(b-c). In particular, the effective  diameter is found to be very close, at all $c$, to the microgel extension $\sigma_{\rm ext}$, displayed in Fig.~\ref{fig:fig2_effpot_interf}(b), that can be estimated by taking opposite edges of the microgel on the interfacial plane~\cite{camerin2019microgels}.
The slight underestimation of $\sigma_{\rm ext}$ as compared to $\sigma_{\rm eff}$ is associated to the fact that effective interaction calculations are also sensitive to the outer dangling chains. This information is partially lost by averaging over the distance of all opposite sites
on the plane of the interface. As expected, the extension of the microgel at the interface decreases as a function of $c$ in agreement with experiments~\cite{camerin2019microgels}, since softer microgels deform more strongly, and hence spread more at the interface.
The corresponding values of $Y$ are reported in Fig.~\ref{fig:fig2_effpot_interf}(c), showing that higher crosslinking leads to stiffer networks, following expectations and in agreement with findings for microgels in bulk~\cite{rovigatti2019connecting}.

\begin{figure}[t!]
\centering
\includegraphics[scale=0.4]{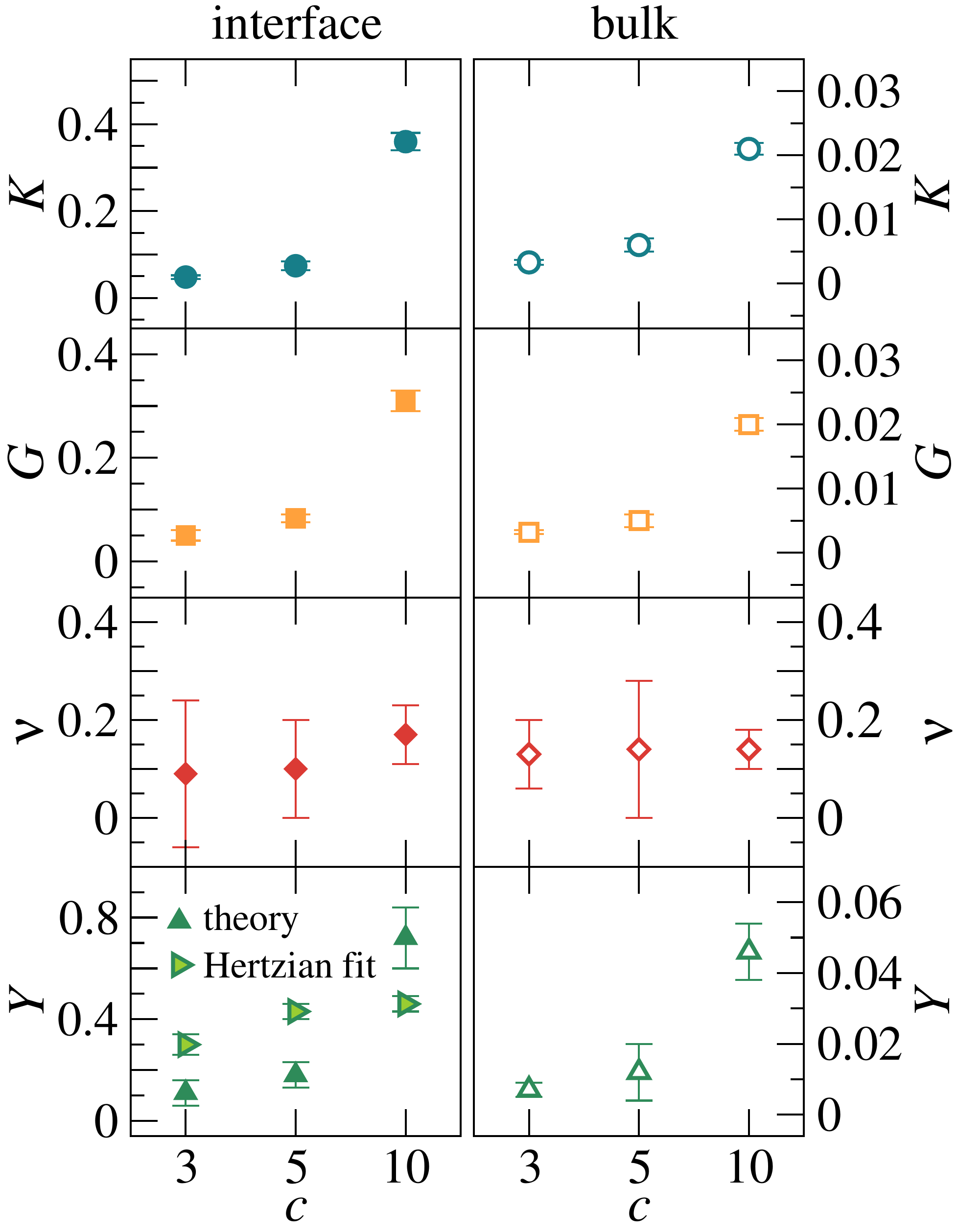}
\caption{\textbf{Elastic moduli at the interface and in the bulk.} Bulk modulus $K$, shear modulus $G$, Poisson's ratio $\nu$ and Young's modulus $Y$ for the same microgel topology at the interface (full symbols) and in bulk (empty symbols) with explicit solvent as a function of $c$.  In the last row, the theoretical results for $Y$ are also compared to the ones obtained from the effective potential fits with the Hertzian model (Eq.~\protect\ref{2dhertz}), also reported in Fig.~\ref{fig:fig2_effpot_interf}(c). $K$, $G$ and $Y$ are in units of $k_BT/\sigma_{\rm m}^3$ to appropriately compare bulk and interface moduli, where the latter are divided by the thickness of the shell at the interface  ($\approx \sigma_{\rm m}$);  $\nu$ is dimensionless. Error bars estimated from the fits of $P(J)$ for $K$ and $P(I)$ for $G$ (see Methods) are propagated in the calculation of $\nu$ and $Y$.}
\label{fig:fig4_elasticmoduli}
\end{figure}

\subsection{Elasticity theory calculations}
The estimate of the Young's modulus extracted from the fit can be compared to the one obtained through the use of elasticity theory in 2D. 
In this framework, one can evaluate the area and shape fluctuations of the microgel on the plane of the interface, writing
the elastic energy $U$ as a function of the two strain invariants of the strain tensor~\cite{doghri2013mechanics,aggarwal2016nonuniform}. In this case, we write $U$ according to the phenomenological Mooney-Rivlin theory, that is known to be valid also beyond the linear elastic regime. Within this theoretical approach, we can calculate all the elastic moduli of a microgel from equilibrium simulations, as previously done for microgels in bulk~\cite{rovigatti2019connecting}. In particular, the moduli refer to the two-dimensional projection of the microgel on the interface, assuming that they are dominated by corona fluctuations. In order to compare with the corresponding bulk properties, we also perform a similar procedure in 3D for the same microgel topologies in the presence of  explicit solvent (see Ref.~\cite{rovigatti2019connecting} for the implicit solvent treatment).
However, bulk and interfacial moduli are naturally given in different units, so we adopt the so-called plane-stress approximation for the 2D moduli. In this way, we assume that the stress normal to the interface is zero~\cite{meyers2008mechanical,hegemann2018elastic}, a legitimate assumption for two-dimensional objects. Hence, we consider the small thickness of the microgels normal to the interface to be roughly comparable to the monomer size ($\approx \sigma_{\rm m}$), and divide the obtained 2D moduli by this length. 
We can finally convert them into the corresponding 3D moduli for very thin three-dimensional objects using the relations reported by Torquato~\cite{Torquato2002random} for plane-stress conditions.  More details on these calculations are provided in the Methods section and in the Supplemental Material.
\begin{figure*}[t!]
\centering
\includegraphics[scale=0.48]{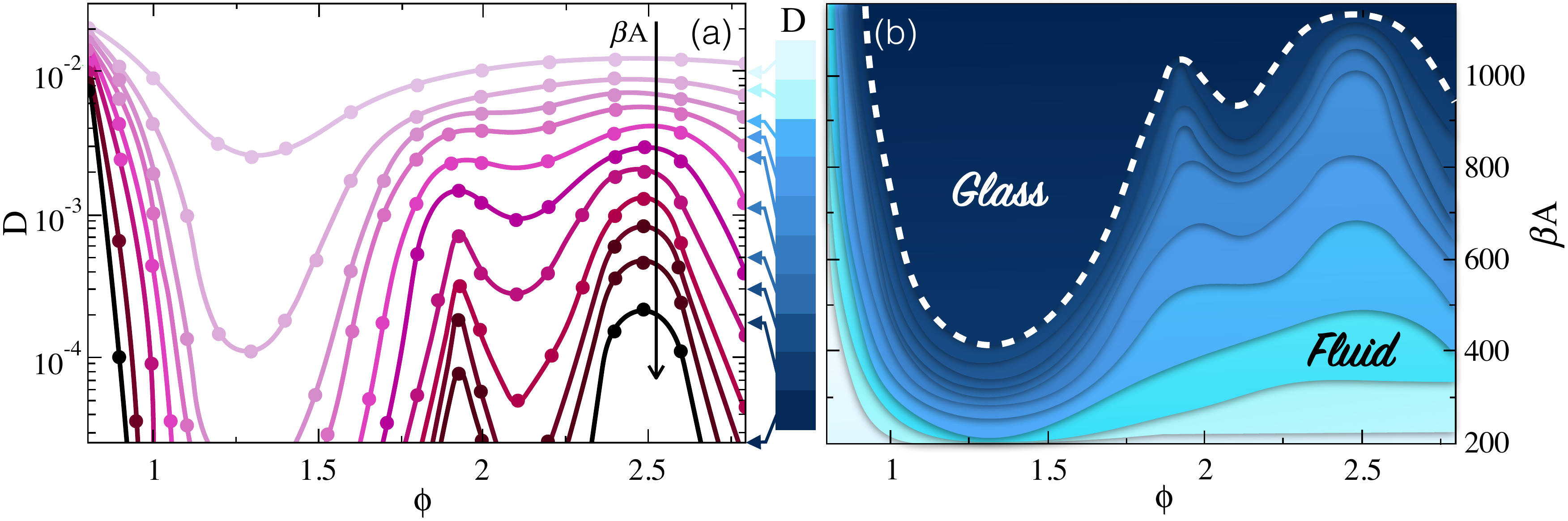}
\caption{\textbf{2D Hertzian phase diagram.} (a) Diffusion coefficient $D$ as a function of the area fraction $\phi$ for different values of the 2D Hertzian strength $A$. From top to bottom, $A$ takes the following values: 226, 340, 409, 453, 566, 680, 793, 906, 974, 1042, 1133 $k_BT$; symbols are simulation data and lines serve as guides to the eye. The lowest reported value of $D$ is taken as the non-ergodicity limit; (b) phase diagram showing $\beta A$ as a function of $\phi$, extracted by taking the iso-$D$ lines from (a). The dashed line signals the onset of the glass region; state-points with the same color-coding have the same value of diffusion coefficient.}
\label{fig:fig4_phasediagram}
\end{figure*}

The resulting elastic moduli are reported in Fig.~\ref{fig:fig4_elasticmoduli} as a function of the crosslinker concentration both for microgels at interfaces (left panels) and in bulk (right panels).
Overall, we observe a monotonic increase of $K, G$ and $Y$ as a function of $c$, while $\nu$ remains nearly constant. We note that the non-linear dependence of $G$ and $Y$ on $c$ is to be expected, since the chains are not Gaussian and the network contains both dangling ends and loops~\cite{lin2019revisiting,kloczkowski2002effect}. These trends are preserved both in the bulk and at the interface. We stress that our two independent estimates of the Young's modulus, namely the one provided by the Mooney-Rivlin theory and that obtained by the 2D Hertzian fitting, also reported in Fig.~\ref{fig:fig4_elasticmoduli}, are consistent with each other. We highlight in this way how the single-particle properties of a microgel at an interface are fully reflected in the multi-particle behavior.
The most striking result of this analysis is the fact that all three moduli at the interface are significantly larger, by approximately one order of magnitude, than their respective bulk counterparts.  As for the Poisson's ratio, even though we find similar values in both cases,
it should be noted that its upper limit in 2D is 1.0 while in 3D is 0.5~\cite{meille2001linear}.
\\
\indent
These findings provide a robust evidence of the reduced flexibility of the microgels at a liquid-liquid interface, an issue that up to now has either been extracted from indirect results or sometimes related to charge effects~\cite{wellert2014inner,kyrey2019influence}. Instead, we directly prove that it is entirely attributable to the presence of the interface, where microgels assume a much more stretched configuration with respect to their standard arrangement in bulk. We are able to establish this link thanks to the relative simplicity of our model, whereby a neutral microgel spontaneously adsorb at the interface without any externally-imposed confinement.
Under these conditions, microgels are much more resistant to deformation. Indeed, the corona is completely extended and restrained at the interface with the polymer chains being much less responsive to external forces than in bulk, while still minimizing the surface tension. We further note that no available experimental results have so far reported the lateral elastic response of the microgels on the interfacial plane but rather the perpendicular one over a solid substrate~\cite{schulte2018probing}. The lateral response is supposed to be the relevant one for the formation of thin microgel layers or for pattern formation on surfaces~\cite{fernandez2018tunable}.

\subsection{Multi-particle dynamical response}
\begin{figure}[t]
\centering
\includegraphics[scale=0.57]{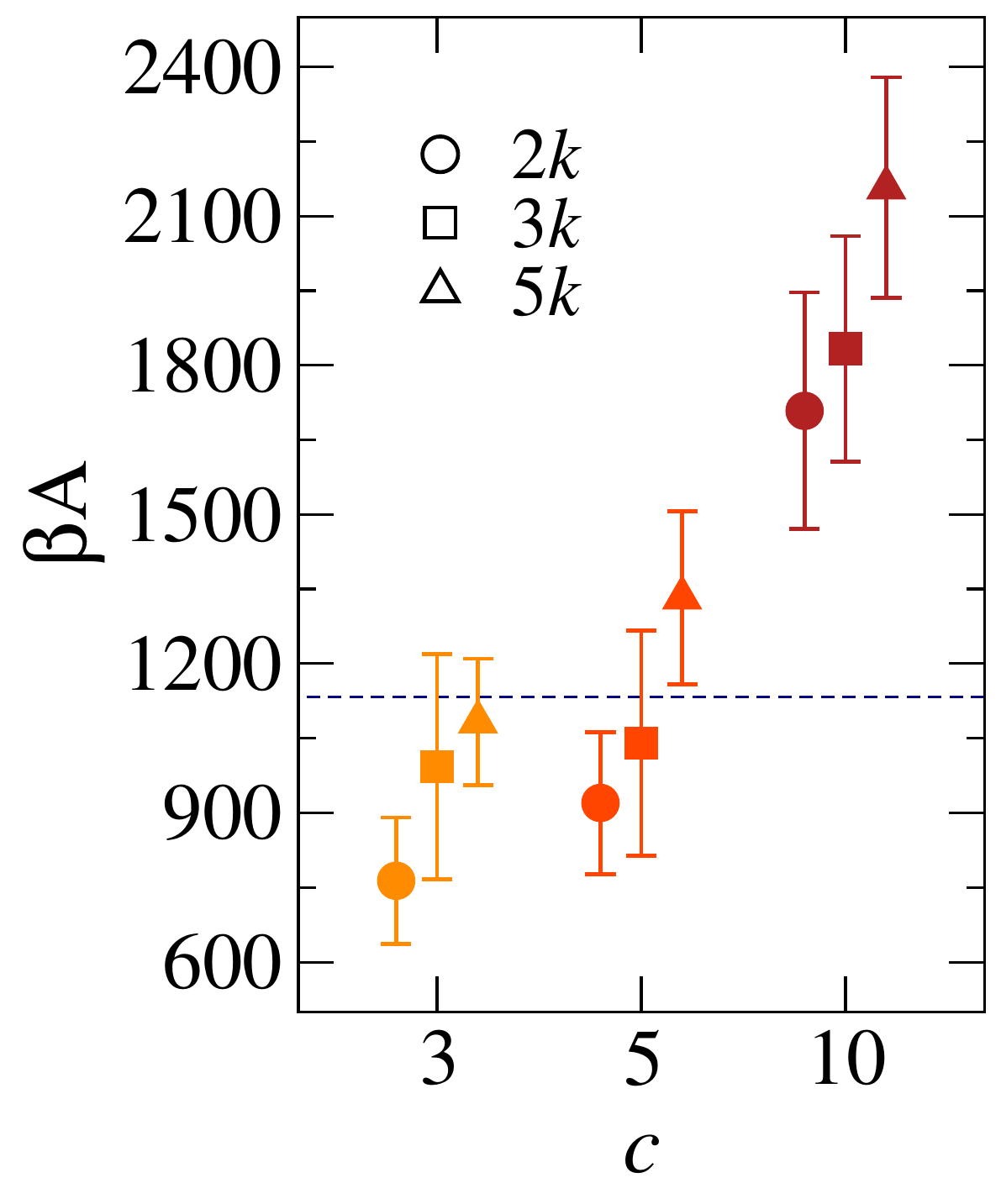}
\caption{\textbf{Dependence of the 2D Hertzian strength $A$ on the crosslinker concentration for microgels of various sizes.} Value of $\beta A$ are extracted \textit{via} the theoretical calculation of $Y$ and $\sigma_{\rm ext}$ for microgels assembled with $N\approx 2000$ (circles), $3000$ (squares) and $5000$ (triangles) monomers for $c=3\%$ (orange), $5\%$ (red) and $10\%$ (dark red). Symbols are slightly displaced on the x-axis to enhance readability. The dashed line indicates the largest value of the Hertzian strength for which a reentrant transition could be observed (see Fig.~\ref{fig:fig4_phasediagram}). Data are averaged over four different topologies for each combination of $N$ and $c$.
}
\label{fig:fig5_sizedependence}
\end{figure}

\indent
The level of coarse-graining adopted up to now has allowed us to describe how the properties of single constituents affect their mutual interactions. Now we go one step further by investigating the 
multi-particle behavior, \textit{i.e.} the condition where many microgel particles interact on the interfacial plane.
To shed light on this aspect,  we simulate a system of particles whose interaction potential is the one we extracted previously, that is the 2D Hertzian potential. In this way, by further coarse-graining our system, we are able to assess for instance the dynamical response of microgels that are adsorbed on the interfacial plane.
\\
\indent
The research on the phase behavior of soft colloids has recently gained much interest: being the archetype potential to describe interactions among elastic particles, the Hertzian phase diagram has been studied both in three~\cite{pamies2009phase,berthier2010increasing} and in two dimensions~\cite{miller2011two,fomin2018phase}. In the latter case, however, the investigations that have been carried out were limited to a change in the value of the exponent of the well-known 3D Hertzian without considering that a variation in the dimensionality of the problem implies a change in the functional form itself. Indeed, the logarithmic correction arising in Eq.~\ref{2dhertz} cannot be properly captured by a simple variation in the Hertzian exponent. 
\\
\indent
We perform Langevin Dynamics simulations of 2D Hertzian particles for different area fractions $\phi$ and varying the strength of the 2D Hertzian $A=\pi Y\sigma_{\rm eff}^2/(2\ln2)$, which corresponds to the $r\rightarrow 0$ limit in Eq.~\ref{2dhertz}. In order to have access to the dynamical response, we avoid crystallization by introducing polydispersity in the system (see Methods). 
It is important to notice that, in our simulations, particles are assumed to have fixed size, differently from bulk conditions where recent simulations and experiments have shown that deswelling plays an important role for concentrated microgel suspensions~\cite{scotti2019deswelling,de2017deswelling,urich2016swelling}.
Instead, there is no reported evidence of deswelling when microgels are compressed at the interface. This is again due to the dominant role of the surface tension which makes adsorbed microgels much less responsive to external stimuli~\cite{harrer2019stimuli,bochenek2019effect}.  In this way, their compression is simply associated with a smooth and monotonic decrease of their interparticle distance, as described experimentally in Ref.~\cite{scheidegger2017compression}.
\\
\indent
Figure~\ref{fig:fig4_phasediagram}(a) reports the self-diffusion coefficients $D$ extracted from the long-time behavior of the mean-squared displacements of the effective microgels for different values of $\phi$ and $A$. We consider the system to fall out-of-equilibrium on the simulation timescale when $D$ decreases by roughly three orders of magnitude with respect to its low-density value. Thus, we assume the system to be arrested for $D\lesssim 2\times 10^{-5}$ (in simulation units).
\\
\indent
Importantly, we reveal the onset of two clear reentrant melting phenomena where the diffusivity, at first, decreases leading to the formation of a glassy system and then it grows again. This increase takes place primarily for $\phi \gtrsim 1.5$ with a local maximum emerging at $\phi \sim 1.9$. For higher densities, after a further slowdown, the system re-fluidifies again acquiring a finite diffusion coefficient. Interestingly, at the new local maximum appearing for $\phi \sim 2.5$, the value of $D$ is even larger than that at the previous maximum.
\\
\indent
Previous works have shown that one can estimate the locus of the glass transition by monitoring the so-called iso-diffusivity (iso-$D$) lines~\cite{zaccarelli2002confirmation,foffi2003structural,gnan2014multiple}, along which $D$ remains constant. Importantly, it has been shown that the iso-$D$ lines always maintain, for not too large values of the probed $D$, the same shape as the ultimate line of arrest. Thus, by extrapolating to the $D\rightarrow 0$ limit, it is possible to locate the glassy region of a system.
By taking a set of different isodiffusivity lines in Fig.~\ref{fig:fig4_phasediagram}(a), we draw the corresponding fluid-glass state diagram for the 2D Hertzian model, shown in Fig.~\ref{fig:fig4_phasediagram}(b). We notice that for the present system a fluid-like region persists at high densities for $\beta A \lesssim 1100$. 
We also stress that similar reentrant features in the dynamics have long been predicted in the three dimensional version of the Hertzian potential~\cite{berthier2010increasing} and in extensive simulations of monomer-resolved single-chain nanoparticles~\cite{verso2016tunable}. This phenomenon is typically linked to the soft nature of the interaction potential that,
in contrast to hard-core ones where the packing of the particles is limited by excluded volume interactions, makes it possible to restore long-time diffusive motion at high densities, thanks to a balance between energetic and entropic contributions, as also observed in simulations of the Gaussian core model~\cite{krekelberg2009anomalous} or of the star polymer potential~\cite{foffi2003structural}.
Nevertheless, reentrant transitions have never been found in experiments of soft~\cite{philippe2018glass} and ultrasoft colloids~\cite{gupta2015validity}. While microgels in bulk conditions do not show high-density liquid states due to their deviations from an ideal Hertzian behavior~\cite{bergman2018new,scheffold2010brushlike}, as also confirmed experimentally~\cite{philippe2018glass,conley2019relationship,scotti2020flow}, those at interfaces stand as optimal candidates for displaying such an intriguing dynamical behavior.
\\
\indent
Crucially, thanks to the knowledge of the functional form of the potential, we can now predict the experimental features of microgels that will most likely show a reentrant behavior. Indeed, since the repulsive Hertzian strength $A$ depends on the Young's modulus and on the effective diameter of the particles, we need to consider microgels whose combined spreading and elastic properties at the interface fall into the predicted reentrant range.  It turns out that we need to focus on microgels with relatively small size, since a reduction of the particle diameter  strongly affects the value of the Hertzian strength, which depends quadratically on it. To be more precise, we perform additional simulations of microgels made of 2000 and 3000 monomers, besides those with $\approx 5000$ monomers. In order to avoid long computational times for the calculation of the effective interactions, we directly determine the Hertzian strengths \textit{via} elasticity theory calculations and by measuring $\sigma_{\rm ext}$ for single particles with different sizes and crosslinker concentrations at the interface. 
\\
\indent
We report the estimated repulsive strengths as a function of $c$ in Fig.~\ref{fig:fig5_sizedependence} and find that soft and small microgels have an Hertzian strength that falls in the range where a reentrant behavior of the dynamics is present, according to the phase diagram in Fig.~\ref{fig:fig4_phasediagram}(b).  We also confirm that the value of the Young's modulus does not exhibit a strong size dependence, especially for $c=3\%$ and $5\%$ (see Supplemental Material), in qualitative agreement to experimental findings on microgels of different sizes~\cite{hashmi2009mechanical,voudouris2013micromechanics,di2015macroscopic}. 
Hence, from this analysis, we conclude that highly crosslinked microgels will always display glassy dynamics at the interface, independently on their size. Instead, low-crosslinked microgels whose Young's modulus at the interface is around $0.1-0.3k_BT/\sigma_{\rm m}^2$ and whose extended size is between $\approx 35-50\sigma_{\rm m}$ are expected to show a reentrant dynamics. Thanks to the mapping performed in previous comparisons with experiments~\cite{ninarello2019modeling,camerin2019microgels}, we are now able to convert these predictions to real values which, for laboratory microgels, correspond to hydrodynamic diameters in bulk $\lesssim 200 $ nm. This value is 
well within the commonly investigated experimental range and offers the additional advantage that capillary effects should be less relevant.
Therefore, adsorbed microgels of small size and low crosslinking ratio constitute a realistic model system to experimentally investigate the presence of a reentrant dynamics, long postulated in the realm of soft colloids.
\\
\indent
It is also instructive to think where this regime can be observed in terms of compression isotherms to which experiments typically refer to. From the present calculations, we estimate that the value of the area fraction is reduced by about a half as compared to the corona-corona contact at low densities. Even though these compressions are not too high~\cite{scheidegger2017compression}, a number of critical issues may emerge and these are ultimately linked to the real-time visualization of the microgels at the interface, which is essential to retrieve dynamical information from the interfacial microgel assembly and thus observe the reentrant melting at high densities. Currently most of the studies are performed \textit{ex-situ} by means of atomic force microscopy (AFM) on silica wafer or similar techniques, from which only structural static information can be extracted. However, the real-time visualization is just one of the option for the experimental verification since other approaches could be devised. For instance, one could imagine to put forward a rheological investigation and analyze the response of the microgel ensemble at different packing fractions. 
Overall, we believe that our predictions will stimulate experimental work to confirm the predicted dynamical behavior for microgels at interfaces.

\section{Conclusions and perspectives}
In summary, in this work we have provided the first numerical estimate of  the two-body effective interaction potential of microgel particles adsorbed at an interface. The complex arrangement of such particles on the interfacial plane is thus rationalized with a simple functional form that reveals that they interact like effective elastic disks with a Young's modulus that increases with the crosslinker concentration. Notably, the values of the elastic moduli at the interface, after appropriate rescaling, are found to be roughly one order of magnitude higher than the one measured in bulk, as also confirmed by elasticity theory calculation of single microgel particles. This can be 
attributed to the dominant effects of the interfacial tension, which controls the response of the polymer network to an external stress, making it much stiffer and more resistant to deformation with respect to the same network in good solvent conditions. This result  has profound consequences on the properties of a generic interfacial assembly of soft colloids, not limited to microgel particles. Indeed, we expect that the reduced mobility of the polymer chains and their enhanced stiffness should be taken into account in the development of novel materials that rely on deformable constituents of any kind. As demonstrated by our results, this effect should be expected at interfacial conditions with large surface tensions, independently of the presence of intrinsic charges in the material or in the fluids. In this respect, our results call for direct experimental verification which could unambiguously shed light on these aspects.
\\
\indent
From a more fundamental perspective, we have clearly demonstrated that the knowledge that is gained on the bulk properties of soft colloids cannot be directly transferred to the interface, which should be considered as a separate case, where particles behave and interact in a different way. Indeed, we have here numerically shown that microgels at interfaces follow the Hertzian predictions as 2D objects 
even at very short separations, well beyond the small-deformation regime.  This is in stark contrast to 
 microgels in bulk, where the validity of the Hertzian model was found to apply only up to interactions of the order of few $k_BT$, corresponding to moderately large center-to-center distances and small deformations. This was due to the internal morphology of the microgel, that imposes multiple length scales to be included in the description of the collective behavior. Instead, at the interface, the behavior is fully dominated by the very extended coronas. We can thus state that microgels do have distinct properties depending on the environment in which they are placed, opening up new avenues for their exploitation. A similar scenario should be expected for any soft particle adsorbed at interfaces with respect to the corresponding behavior in bulk conditions.
\\
\indent
The extensive analysis of the multi-particle dynamics has further evidenced the emergence of reentrant dynamics, where the system behaves as an ergodic fluid up to very large densities, well above individual particles contact, sometimes loosely called jamming. Experimentally, small (nano-sized) soft microgels appear to be the ideal candidates to verify our theoretical predictions, as indicated by the values of the Young's modulus and of the interfacial extension at which the reentrance is observed. In addition, small colloids are the least likely to experience capillary attractions at the interface, and hence will behave more similarly to the ones we simulated.
\\
\indent
It will be important in the future to extend this study to crowded configurations to investigate the validity of the present results at considerably high packing fractions where additional mechanisms, like faceting or interpenetration, may become relevant. Under these conditions, many-body effects should also play a prominent role. In our current treatment, we cannot quantify the influence of many-body interactions due to severe computational limitations, since a huge number of particles should be used.
Our approach provides, in this respect, a first step towards a comprehensive description of microgel interactions at a microscopic level. 
Similar considerations should be extended to microgels in bulk conditions, for which high-density states still require appropriate theoretical assessment. The analysis can be further broadened to other microgel topologies that have recently gained considerable attention, such as hollow~\cite{geisel2015hollow}, ultra-low-crosslinked~\cite{scotti2019exploring} or anisotropic ones~\cite{nickel2019anisotropic}.
\\
\indent
All in all, our study opens the way for the investigation of microgels at the interface as a simple realization of 2D elastic particles. We expect that the evidence reported here will have important consequences on the study of two-dimensional elastic objects at the fundamental level~\cite{peng2017diffusive,vialetto2019photoswitchable} and for the clever design of composite materials~\cite{maza2018layer,serpe2003layer,murray2019microgels}.

\section{Methods}
\subsection{Modeling and simulation details}
Microgels are assembled
starting from an ensemble of two and four-folded patchy particles in a spherical cavity.  Bivalent and tetravalent particles mimic, respectively, NIPAM (N-Isopropylacrylamide) monomers and BIS (N,N'-methylenebisacrylamide) crosslinkers in a chemical synthesis~\cite{gnan2017silico}. Microgel assembly is performed with the oxDNA simulation package~\cite{oxDNA_edge}. The topology of the polymer network, whose monomers have diameter $\sigma_{\rm m}$ (which also sets the unit of length) is then fixed by means of a classical bead-spring model~\cite{kremer1990dynamics} that amounts to the Weeks-Chandler-Andersen (WCA) potential for non-bonded monomers, and a sum of the WCA and the Finitely Extensible Nonlinear Elastic (FENE) potentials for bonded ones:
\begin{equation}\label{wca}
V_{\rm WCA}(r)=\begin{cases}
    4\epsilon\left[\left(\frac{\sigma_{\rm m}}{r}\right)^{12}-\left(\frac{\sigma_{\rm m}}{r}\right)^{6}\right] + \epsilon  & \text{if $r \le 2^{\frac{1}{6}}\sigma_{\rm m}$}\\
    0 &  \text{otherwise}
  \end{cases};
\end{equation}
\begin{equation}
V_{\rm FENE}(r)=-\epsilon k_FR_0^2\ln\left[1-\left(\frac{r}{R_0\sigma_{\rm m}}\right)^2\right]     \text{ if $r < R_0\sigma_{\rm m}$}
\end{equation}
with $k_F=15$ a dimensionless spring constant and $R_0=1.5$ the maximum extension of the bond. The method ensures that each designed microgel is made by a disordered cross-linked polymer network with a core-corona structure as in real microgels. Furthermore, the adoption of a designing force acting on the cross-linkers during the network assembly improves considerably the agreement between the numerical and experimental form factors as described in Ref.~\cite{ninarello2019modeling}. Due to the fact that each microgel is assembled independently, the topology between different particles may slightly vary in terms of internal connectivity and density profiles, while maintaining the same polymer chains distribution and the same macroscopic features such as the presence of a core-corona structure or a density profile that slowly decrease towards the outer shells. We stress the importance to have realistic topologies that can closely match experimental data~\cite{camerin2019microgels}, since the inner structure of the microgel particle influence both the elastic properties and the effective interactions.
\\
In the present work, we employ microgels with $\approx5000$ monomers and three different molar crosslinker concentrations $c$, namely  $3\%, 5\%$ and $10\%$. The radius of the confining sphere into which microgels are assembled is set to $25\sigma_{\rm m}$. 
We also evaluate the elasticity of smaller microgels with $\approx 2000$ and $3000$ particles assembled in the same way and maintaining the same internal monomer density. The effective interactions are assessed for a second microgel topology for $c=5\%$ (see Supplemental Material), while the analysis of the elastic properties as a function of the microgel size are studied over four independent microgel topologies.
\\
\indent
According to previous works~\cite{camerin2018modelling,camerin2019microgels}, the solvent is modeled within the Dissipative Particle Dynamics (DPD) framework~\cite{groot1997dissipative}. The interactions are described by three forces, conservative $\vec{F}^C_{ij}$, dissipative $\vec{F}^D_{ij}$ and random $\vec{F}^R_{ij}$, of form:
\begin{equation}\label{dpd_fc}
\vec{F}^C_{ij}=\begin{cases}
    a(1-r_{ij}/r_c)\hat{r}_{ij} & \text{if $r_{ij}<r_c$},\\
    0 & \text{otherwise}
  \end{cases}
\end{equation}
\begin{equation}
\vec{F}^D_{ij}=-\xi w^D(r_{ij})(\hat{r}_{ij} \cdot \vec{v}_{ij}) \hat{r}_{ij}
\end{equation}
\begin{equation}
\vec{F}^R_{ij}=\sigma_R w^R(r_{ij}) \theta_{ij} (\Delta t)^{-1/2}\hat{r}_{ij}
\end{equation}
where $\vec{r}_{ij}=\vec{r}_i-\vec{r}_j$, with $\vec{r}_{i}$ the position of particle $i$, $r_{ij} = |\vec{r}_{ij}|$, $\hat{r}_{ij} = \vec{r}_{ij} / r_{ij}$, $r_c$ the cutoff radius, $\vec{v}_{ij}=\vec{v}_i-\vec{v}_j$ with $\vec{v}_{i}$ the velocity of particle $i$, $a$ is the maximum repulsion between two particles, $\theta_{ij}$ is a Gaussian random number with zero mean and unit variance and $\xi$ is the friction coefficient~\cite{groot1997dissipative}. To ensure that Boltzmann equilibrium is reached, $w^D(r_{ij})=[w^R(r_{ij})]^2$ and $\sigma^2_R=2\xi k_BT$ with $k_B$ the Boltzmann constant and $T$ the temperature.
\\
\indent
In order to reproduce a water/hexane (w/h) interface, we choose $a_{\rm ww}=a_{\rm hh}=8.8$, $a_{\rm hw}=31.1$. While in principle it is possible to change these parameters in order to obtain a different surface tension, no significant difference is expected in the microgel-microgel interaction potential, given that the distribution of microgel monomers remains unaltered~\cite{camerin2019microgels}. Instead, for the monomer-solvent interactions we choose $a_{\rm mw}=4.5$ and $a_{\rm mh}=5.0$.  The cut-off radius is always set to be $r_c=1.9 \sigma_{\rm m}$ and the reduced solvent density $\rho_{\rm DPD}=4.5$~\cite{camerin2019microgels}. Depending on the microgel size, up to $\approx 750000$ solvent particles are inserted in the simulation box. To analyze the elastic properties of microgels in bulk, we also run bulk simulations with explicit solvent. In there, the solvent-solvent parameters are not varied with respect to interfacial simulations, while $a_{\rm ms}=1.0$, ensuring good solvent conditions. A more detailed discussion on how these parameters have been determined can be found in Ref.~\cite{camerin2019microgels}.
\\
\indent
Simulations are carried out using the LAMMPS simulation package~\cite{plimpton1995fast}. The equations of motion are integrated with a velocity-Verlet algorithm. The reduced temperature $T^*=k_BT/\epsilon$ is always set to $1.0$  \textit{via} the DPD thermostat~\cite{camerin2018modelling}.
Length, mass and energy are  given in units of $\sigma_{\rm m}$, $m$, $\epsilon$, respectively. DPD repulsion parameters $a$ are in units of $\epsilon/\sigma_{\rm m}$. 
\\
\indent
The phase behavior of the 2D Hertzian potential is assessed by means of molecular dynamics simulations in two dimensions with 5000 particles of unit mass $m$ and diameter $\sigma_{\rm eff}$, interacting \textit{via} Eq.~\ref{2dhertz}. We use $\sigma_{\rm eff}$ as the unit of length, so that the area fraction is defined as $\phi=\frac{\pi}{4}\langle\sigma_{\rm eff}^2\rangle\rho$, with $\rho$ the number density. We fix $k_BT=1$, which defines the unit of energy, \textit{via} a Langevin thermostat; time is in units $\sqrt{m\sigma_{\rm eff}^2/k_BT}$.
To avoid crystallization, we set the polydispersity to $p=0.2$. We analyze a range of $\phi$ from $0.8$ to $2.8$ for a Hertzian strength, defined as $A=\pi Y\sigma_{\rm eff}^2/(2\ln2)$, that goes from $220$ to $1150 k_BT$. We note that $A$ has units of energy over length squared, meaning that it changes value depending on the units of measurement used ($\sigma_{\rm m}$ and $\sigma_{\rm eff}$ for the monomer-resolved system and the coarse-grained systems, respectively).
For all $\phi$ and $A$, we monitor the presence of a liquid-like disordered structure by calculating the radial distribution function (see Supplemental Material).
\\
To determine the glass region in the 2D phase diagram, we run simulations for $\sim 2\times 10^7$ timesteps and we calculate the mean-squared displacement $\left\langle \Delta r^2\right\rangle$ of the particles, extracting the long-time self-diffusion coefficient $D$: 
\begin{equation}
D= \lim_{t\rightarrow \infty}\frac{\left\langle \Delta r^2\right\rangle}{4t}
\end{equation}
where $t$ is the simulation time. Since we are only interested in providing a state diagram assessment, we do not perform an extensive characterization of the glassy dynamics of the system and we just monitor the onset of non-ergodicity within the timescale of our simulations~\cite{berthier2010increasing}.  We  attribute this condition to state points where we find $D\lesssim 2.5\times 10^{-5}$, roughly three orders of magnitude lower than the corresponding low-density value. Under these conditions, the system has become so slow that aging is present within our simulation time window.
\\

\subsection{Calculation of the effective interaction potential}
The two-body effective potential between the microgels at the interface is evaluated by means of the Umbrella Sampling technique in explicit solvent~\cite{blaakaccurate,likos2001effective,gnan2014casimir}. This method allows to  uniformly sample all distances between the centers of mass of the microgels by adding a harmonic potential between them. 
For each sampled window $i$, we evaluate the probability distribution $P(r,\Delta_i)$ of finding the microgels' centers of mass at distance $r$ given the equilibrium length of the spring $\Delta_i$. The final probability for the entire range of explored distances is obtained by first removing the contribution of the bias potential and by subsequently merging $P(r,\Delta_i)$ into $P(r)$ for all the  windows \textit{via} a least-square method. Finally, the potential of mean force $V_{\rm eff}$ is retrieved knowing that 

\begin{equation}
V_{\rm eff}=-k_BT\ln(P(r))+C,
\end{equation} 

where $C$ is such that $V_{\rm eff}(r\rightarrow\infty)=0$.
The major drawback of studying the effective interactions with an explicit solvent model is the computational cost required to carry out the simulations: about two months on about 80 CPU cores are needed to investigate a range from $20$ to $30\Delta_i$.

\subsection{Assessment of the elastic moduli}
Following Ref.~\cite{treloar1976mechanics}, the elastic energy $U$ of a two-dimensional object can be written as a function of the invariants $J$ and $I$ of the strain tensor as
\begin{equation}
U(J, I)=U_0+W(J)+W(I)
\end{equation}
where $U_0$ is the energy of a reference configuration that is taken as the average ellipse adopted by equilibrium configurations of the microgels at the interface. Its semi-axes $s_1$ and $s_2$ are obtained by the gyration tensor built \textit{via} the two-dimensional convex hull on the plane of the interface. We approximate $W$ with the corresponding potentials of mean force
\begin{equation}\label{pmf}
W(X)=-k_BT\ln P(X)+D
\end{equation}
with $X=J, I$. $P(X)$ is the respective probability distribution and $D$ an arbitrary constant.
These potentials can then be fitted to appropriate functions
\begin{equation}\label{function}
f(X; M_X, X_0, \gamma, C)=M_X(X-X_0)^\gamma+C
\end{equation}
with $\gamma=2$ when $X=J$ and $\gamma=1$ when $X=I$, to obtain $M_J$ and $M_I$. In the Supplemental Material, we report, as an example for $c=10\%$, the simulation outcomes and their relative fits both for the microgel at the interface and in bulk. 
The elastic moduli are then readily obtained as
\begin{equation}
K=\frac{2M_J}{S}
\end{equation}
\begin{equation}
G=\frac{2M_{I}}{S}
\end{equation}
with $S=\pi s_1 s_2$. $Y$ and $\nu$ only depend on $K$ and $G$ as~\cite{meille2001linear}
\begin{equation}
\nu=\frac{K-G}{K+G}
\end{equation}
\begin{equation}
Y=\frac{4KG}{K+G}
\end{equation}
Similar expressions can be derived for the 3D case and can be found, for instance, in Ref.~\cite{rovigatti2019connecting} (see also below).

The particular choice of $W$ as a function of $J$ and $I$ depends on the specific elastic model employed. Here, we considered the Mooney-Rivlin model
for which the elastic energy reads~\cite{doghri2013mechanics,aggarwal2016nonuniform}

\begin{equation}
U(J, I)=U_0+\frac{S}{2}\left[K(J-1)^2+G(I-2)\right].
\end{equation}

We have further checked that the obtained results do not crucially depend on the specific form of the employed $W$. To this aim, we also employed the linear elastic model (Hookean)~\cite{sadd2009elasticity,doghri2013mechanics} and the Saint-Venant-Kirchhoff model~\cite{athanasopoulou2017phase,riest2015elasticity}, finding results for the moduli, particularly the Young's modulus, that are very close to the ones presented in the main text. They display the same increase with respect to the bulk model and a similar monotonic increase with $c$. 

To convert the 2D moduli into 3D ones, we consider that for two-dimensional objects the stress normal to the interfacial plane is zero. Under these conditions, there exist relations to convert 2D moduli into 3D ones, by assuming that the 2D object has a given (small) thickness $h$:
\begin{eqnarray}
 G^{(3)}&=&G^{(2)}/h \\
  Y^{(3)}&=&Y^{(2)}/h\\ 
  K^{(3)}&=&\frac{4 G^{(2)}K^{(2)}}{3h(3G^{(2)}-K^{(2)})},
 \end{eqnarray}
where $X^{(3)}$ indicates the converted 3D moduli (in units of $k_BT/\sigma_{\rm m}^3$) from the 2D results $X^{(2)}$ (in units of $k_BT/\sigma_{\rm m}^2$) with $X=G, Y, K$. Also, we have that~\cite{Torquato2002random}
\begin{equation}
\nu^{(3)}=\nu^{(2)}.
\end{equation}

In our case, we consider $h$ to be roughly equal to the monomer size, $\sigma_{\rm m}$, as in the outer shells chains do not pile up, but remain confined to the interfacial plane, providing the dominant contribution to the elastic response of the microgels. Furthermore, as reported in Ref.~\cite{camerin2019microgels} based on AFM studies, the realistic width of a microgel corona is below 7 nm, that is a fully compatible size to the one we extract for a single \textit{in silico} microgel monomer by comparing the form factors of numerical and laboratory microgels~\cite{ninarello2019modeling}.

\section*{Acknowledgements}
We thank M. A. Fern{\'a}ndez-Rodr{\'\i}guez, F. Grillo, L. Isa, J. Vialetto and P. Ziherl for valuable discussions. We acknowledge financial support from the European Research Council (ERC Consolidator Grant 681597, MIMIC), from the European Union's Horizon 2020 research and innovation programme (Grant 731019, EUSMI), from MIUR (FARE project R16XLE2X3L, SOFTART) and from Regione Lazio, through L.R. 13/08 (Progetto Gruppo di Ricerca GELARTE, n.prot.85-2017-15290). The authors gratefully acknowledge the computing time granted by EUSMI on the supercomputer JURECA at the Jülich Supercomputing Centre (JSC), and by CINECA under the ISCRA initiative.


\clearpage
\newpage
\onecolumngrid
\begin{center}
\large
\textbf{Microgels at interfaces behave as 2D elastic particles featuring reentrant dynamics\\ \bigskip Supplemental Material}

\normalsize
\bigskip
Fabrizio Camerin\textsuperscript{ 1, 2}, Nicoletta Gnan\textsuperscript{ 1, 3}, Jos\'e Ruiz-Franco\textsuperscript{ 3, 1}, Andrea \\ Ninarello\textsuperscript{ 1, 3}, Lorenzo Rovigatti\textsuperscript{ 3, 1}, Emanuela Zaccarelli\textsuperscript{ 1, 3}\\
\medskip
\small
\textit{%
\textsuperscript{1}CNR Institute of Complex Systems, Uos Sapienza, Piazzale Aldo Moro 2, 00185 Roma, Italy\\
\textsuperscript{2}Department of Basic and Applied Sciences for Engineering,\\ Sapienza University of Rome, via Antonio Scarpa 14, 00161 Roma, Italy\\
\textsuperscript{3}Department of Physics, Sapienza University of Rome, Piazzale Aldo Moro 2, 00185 Roma, Italy
}

\end{center}

\normalsize
\bigskip
\twocolumngrid
\renewcommand{\thefigure}{S\arabic{figure}}\setcounter{figure}{0}

\section{Analysis of a different topology}
\begin{figure}[h!]
\centering
\includegraphics[scale=0.33]{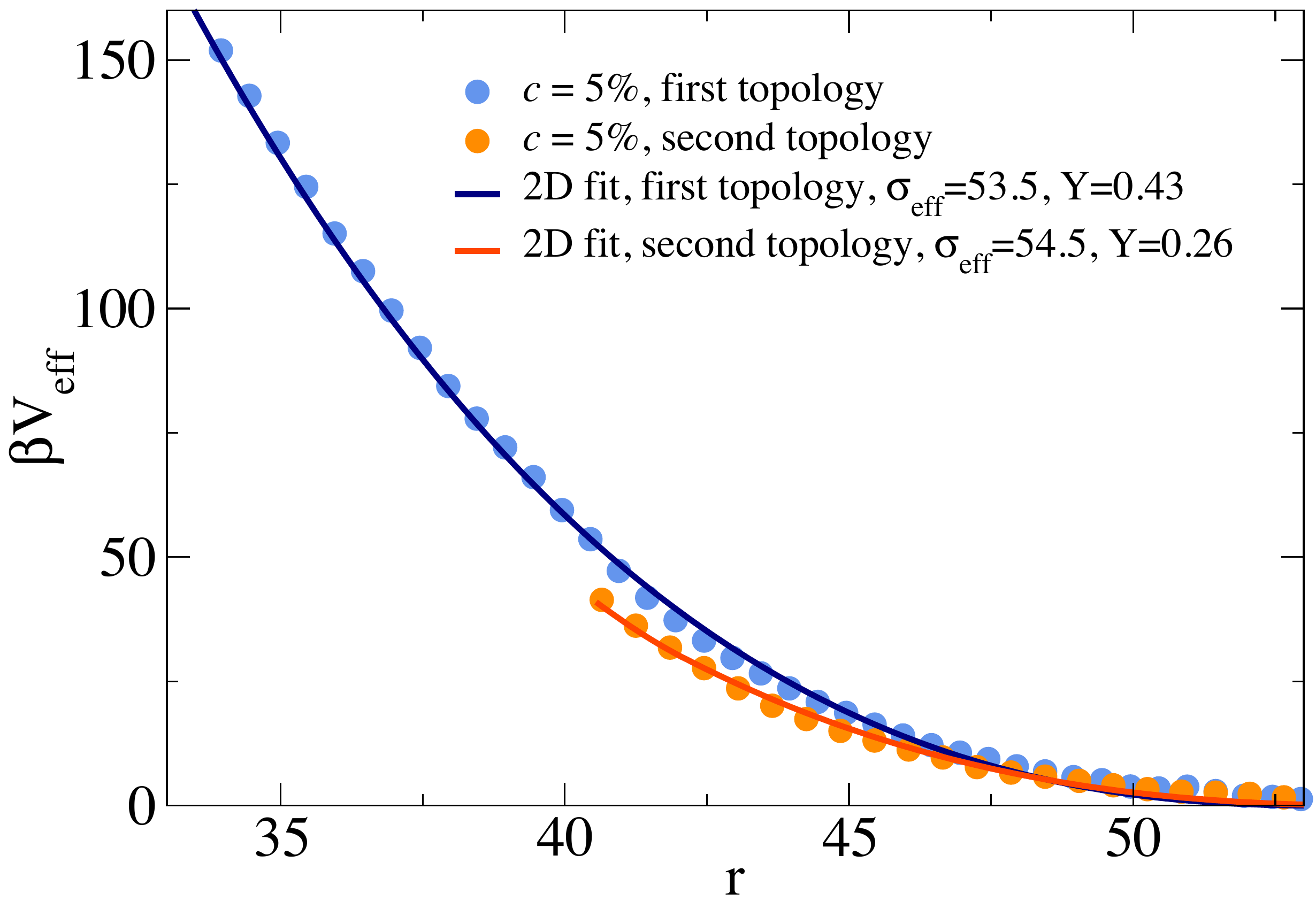}
\caption{\small \textbf{Effective potential for $c=5\%$.} Symbols are simulation results while the full line are fits with the 2D Hertzian potential (Eq. 1 of the main text).}
\label{fig:secondtop}
\end{figure}

Due to the high computational cost of the calculations of the effective potential, we limit the analysis in the main text to a single microgel topology. However, we checked that the presented results are robust in their main conclusions by analyzing a different microgel topology for the case $c=5\%$.
We report in Fig.~\ref{fig:secondtop} the interaction potential for such a different topology, limiting the analysis, again due to the high computational cost, to small compression regimes. Also in this case, we confirm the validity of the fitting procedure with the bidimensional Hertzian potential (Eq. 2 in the main text). The extracted fitting parameters are $\sigma_{\rm eff}=54.5 \sigma_{\rm m}$ and $Y=0.26 k_BT/\sigma_{\rm m}^2$, in good qualitative agreement with the ones of the microgel presented in the main text.

\section{Further considerations on the effective potential}
\begin{figure}[b!]
\centering
\includegraphics[scale=0.31]{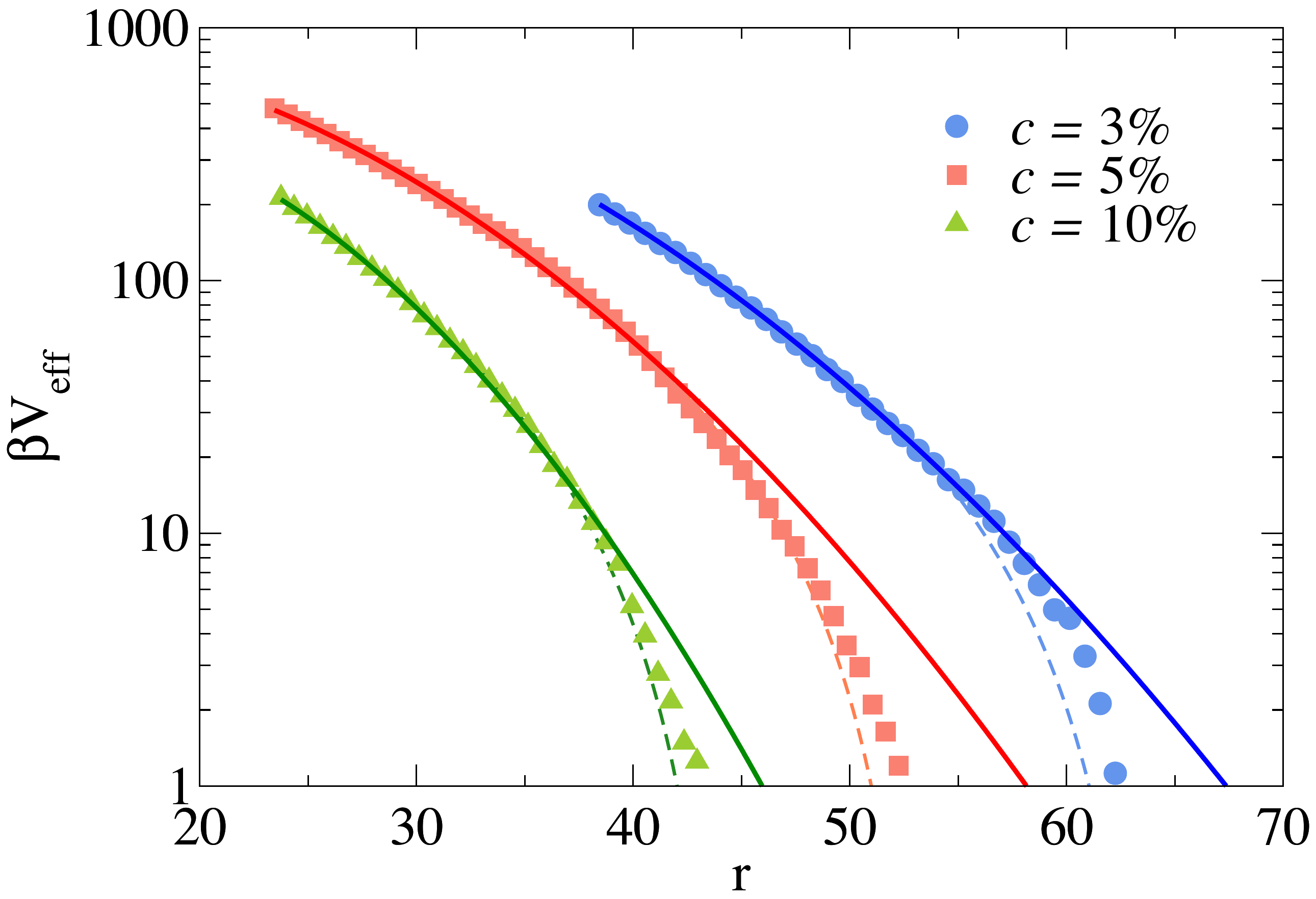}
\caption{\small \textbf{Effective potentials for microgels at an interface and gaussian fit.} Symbols are simulation results, full lines are fits according to Eq.~\ref{gauss} and dashed lines are fits according to Eq. 1 of the main text (2D Hertzian fits).
}
\label{fig:gaussianfit}
\end{figure}
It is interesting to compare the effective potential between microgels at interfaces to a gaussian functional form, that was used to describe brush-coated spherical nanoparticles in bulk~\cite{verso2011interactions,cerda2003pair} and at an interface~\cite{schwenke2014conformations}. Such a simple model can be written as
\begin{equation}\label{gauss}
V_{\rm gaussian}(r)=b \exp(-d(r-e)^2)
\end{equation}
where $b$, $d$, $e$ are fitting parameters and $r$ is the distance between the centers of mass of the particles. From a structural perspective, the conformation that microgels retain at interfaces may resemble the one of such particles, given the presence of extended, flexible polymer chains surrounding a more compact core. For polymer brushes, there exists a scaling theory for the fitting parameters $b$ and $d$~\cite{cerda2003pair}. Surprisingly, the functional form in Eq.~\ref{gauss} was found to describe the calculated interactions for these systems quite well. Nonetheless, there should not be any physical reason for this framework to be applicable to our system. In addition, the gaussian fit does not account for any deformation of the polymer at the interface, being developed for 3D bulk systems. 
\begin{table}[H]
\centering
\begin{tabular}{|c||c|c|c|c}
\cline{1-4}
\textit{c}(\%) & \textit{b}   & \textit{d}     & \textit{e}    &  \\ \cline{1-4}
3               & 910 & 0.002 & 12.7 &  \\ \cline{1-4}
5               & 871 & 0.003 & 8.5  &  \\ \cline{1-4}
10              & 462 & 0.005 & 18.0 &  \\ \cline{1-4}
\end{tabular}
\caption{\small \textbf{Gaussian fit.} Fitting parameters according to Eq.~\ref{gauss}.}
\label{gaussianparam}
\end{table}

\begin{figure}[h]
\centering
\includegraphics[scale=0.3]{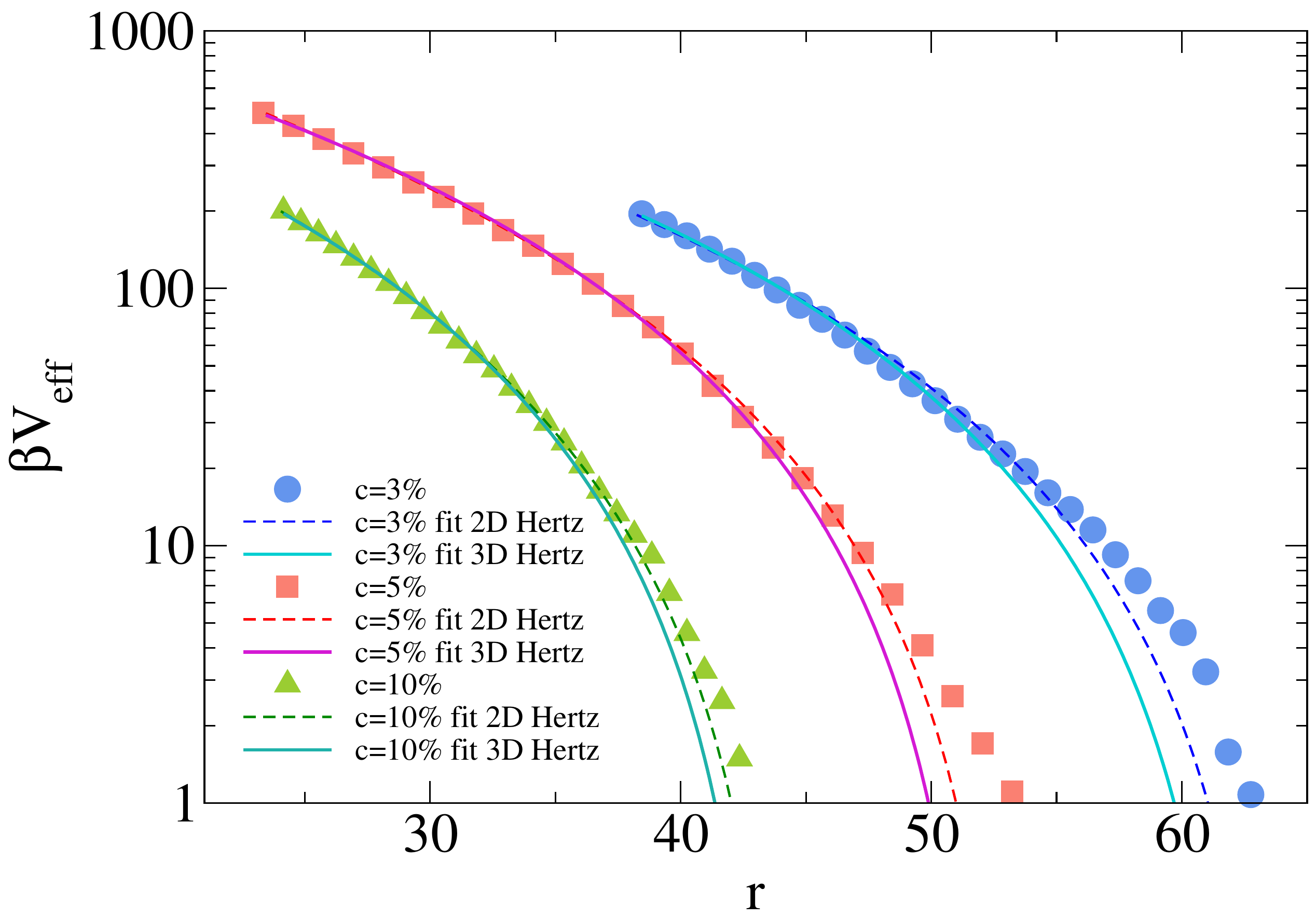}
\caption{\small \textbf{Comparison of different fits to the effective potentials.} For the three values of crosslinker analyzed, we compare the fits of the numerical data (symbols) by using a 2D Hertzian as used in the manuscript (dashed lines) and  a 3D Hertzian potential (solid lines). 
}
\label{fig:effpot}
\end{figure}

We report in Fig.~\ref{fig:gaussianfit} our calculated potentials and the corresponding fits with Eq.~\ref{gauss}  and with the 2D Hertzian model described in the main text. We find that the latter agrees much better with data also at large distances between the microgels, while the gaussian form fails in this regime. Although this is the region in which data are most affected by statistical noise being the probed energy of the order of a few $k_BT$s, the gaussian fit would give rise to a potential which tends  to zero at distances that are clearly non-compatible with the dimensions of the microgel particles analyzed here (see also the comparison with $\sigma_{\rm ext}$ in Fig. 2 of the main text).  
While we could think of operating the gaussian fit in a reduced region of distances, i.e. only for short ones,  it is important to stress that the parameters that we would extract from such fits cannot be related to any physical feature of our system. For the sake of completeness, the fitting parameters for the gaussian functional form are reported in Table~\ref{gaussianparam}, where it is evident that in the case of parameter $e$,  we cannot even identify a clear trend as a function of the crosslinker concentration. 
\\
\\
Given the quasi-2D nature of the adsorbed microgels, it is also interesting to compare the simulation data to the 3D version of the Hertzian potential that reads as
\begin{equation}
V_H^{3D}(r) = U_0 (1-r/\sigma)^{5/2} \theta(1-r/\sigma)
\label{eq:3D}
\end{equation}
where the prefactor, \textit{i.e.} the repulsion strength, $U_0= \frac{2 Y\sigma^3}{15 (1-\nu^2)}$ specifically depends on the size of the particle $\sigma$ and on the zero-stress elastic properties of the single particle, the Young's modulus $Y$ and the Poisson's ratio $\nu$.
\\
We thus fit the effective potentials with Eq.~\ref{eq:3D} using as free parameters $U_0$ and $\sigma$. The resulting values, shown in Table~\ref{table:fitting}, can be compared to $U_0^{th}=\frac{2 Y_{th}\sigma_{ext}^3}{15 (1-\nu_{th}^2)}$ where $Y_{th}$ and $\nu_{th}$ are the  moduli extracted from the Mooney-Rivlin theory, while $\sigma_{ext}$ is the extension of the microgel on the plane of the interface, as defined in the manuscript.
\\
In Fig.~\ref{fig:effpot} we observe not only that the quality of the fits is worse with respect to the 2D model, but, most importantly, we find that, while the value of $\sigma$ remains plausible, there is a strong discrepancy between the estimated $U_0$ and the corresponding theoretical prediction $U_0^{th}$. These results clearly indicate the inadequacy of the 3D Hertzian potential in describing the data. 
\begin{table}[h]
\centering
\small
\begin{tabular}{|c||c|c||c|c|c|c|}
\hline
$c$& $U_0$ & $\sigma$ & $Y_{th}$ & $\nu_{th}$ &$\sigma_{ext}$& $U_0^{th}$ \\ \hline
 3   &2065& 62.6 &0.11&     0.09        &  62.3 &  3575 \\ \hline
 5   &2078  & 52.4 &0.18&     0.1     &     52.5  & 3507   \\ \hline
10 & 1454 &43.7 &0.72&    0.17  &   40.3   &  6470 \\ \hline
\end{tabular}
\caption{\small \textbf{Comparison of 3D Hertzian fit parameters to theoretical values} The 3D Hertzian potential $U_0$ and $\sigma$ are compared with the corresponding strength estimated from theory $U_0^{th}$ and with the calculated extension $\sigma_{ext}$.}
\label{table:fitting}
\end{table}
\\
We thus conclude that the 2D Hertzian description presented in the main text is the most appropriate to treat the effective interactions between microgel particles at an interface.

\section{Elastic moduli}
\begin{figure}[t]
\centering
\includegraphics[scale=0.5]{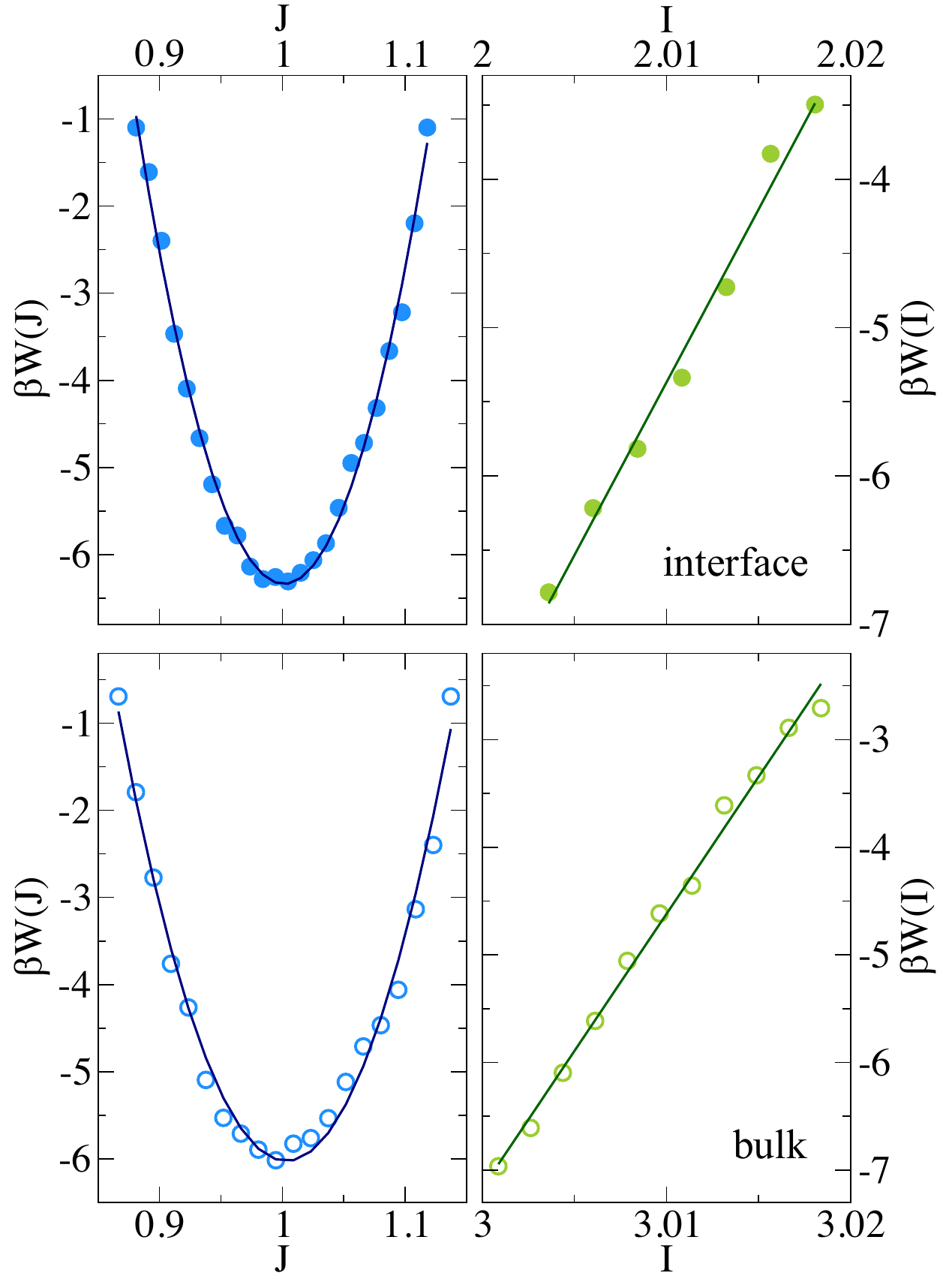}
\caption{\small \textbf{Potential of mean force $W(J)$ and $W(I)$ for a microgel at the interface and in bulk for $c=10\%$.} Symbols are simulation data (interface, upper panels; bulk, lower panels) and full lines are fits according to Eq.~\ref{function}, as described in the main text.}
\label{fig:elasticfit}
\end{figure}
\begin{figure}[t]
\centering
\includegraphics[scale=0.45]{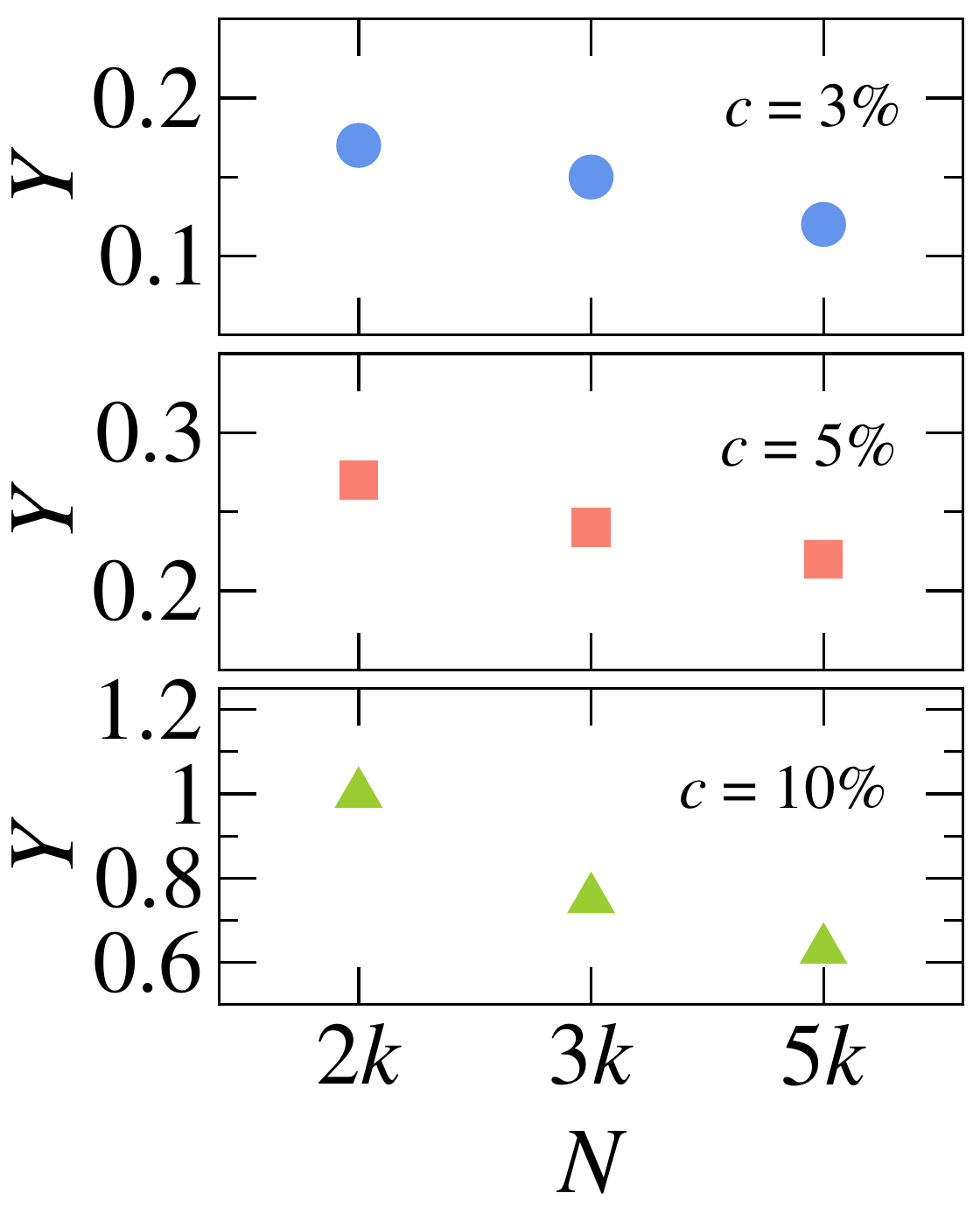}
\caption{\small \textbf{Young modulus for different crosslinker concentrations and microgel sizes.} Dependence of the Young modulus as a function of $c=3\%,5\%,10\%$ for $N=2000,3000,5000$ microgel monomers at the interface. Data are in units of $k_BT/\sigma_m^2$. }
\label{fig:youngmodulus}
\end{figure}
As also explained in the Methods section of the main text, the elastic energy of the microgel can be written as a function of a reference configuration energy and of $W(X)$ with $X=J, I$, being $J$ and $I$ the invariants of the strain tensor~\cite{doghri2013mechanics,aggarwal2016nonuniform}. $W$ can be approximated with the potentials of mean force
\begin{equation}\label{pmf}
W(X)=-k_BT\ln P(X)+C
\end{equation}
with $P(X)$ the respective probability distribution and $C$ an arbitrary constant. This can be fitted to functions of form
\begin{equation}\label{function}
f(X; M_X, X_0, \gamma, C)=M_X(X-X_0)^\gamma+C
\end{equation}
with $\gamma=2$ when $X=J$ and $\gamma=1$ when $X=I$, to obtain $M_J$ and $M_I$. Similar considerations apply to the 3D case, for microgels in bulk~\cite{rovigatti2019connecting}. In Fig.~\ref{fig:elasticfit}, we report, as an example for $c=10\%$, the simulation outcomes and their relative fit both for the microgel at the interface and in bulk. From these, all the elastic moduli are readily obtained.
\\
\indent
In Fig.~\ref{fig:youngmodulus}, we show the dependence of the Young modulus as a function of the size of the microgel. Except for the smallest microgels with $c=10\%$, we observe only a slight dependence on the size of the microgel at fixed crosslinker concentration. This is in qualitative agreement with experimental findings~\cite{di2015macroscopic}.

\section{Multi-particle dynamical response}
\begin{figure}[ht!]
\centering
\includegraphics[scale=0.45]{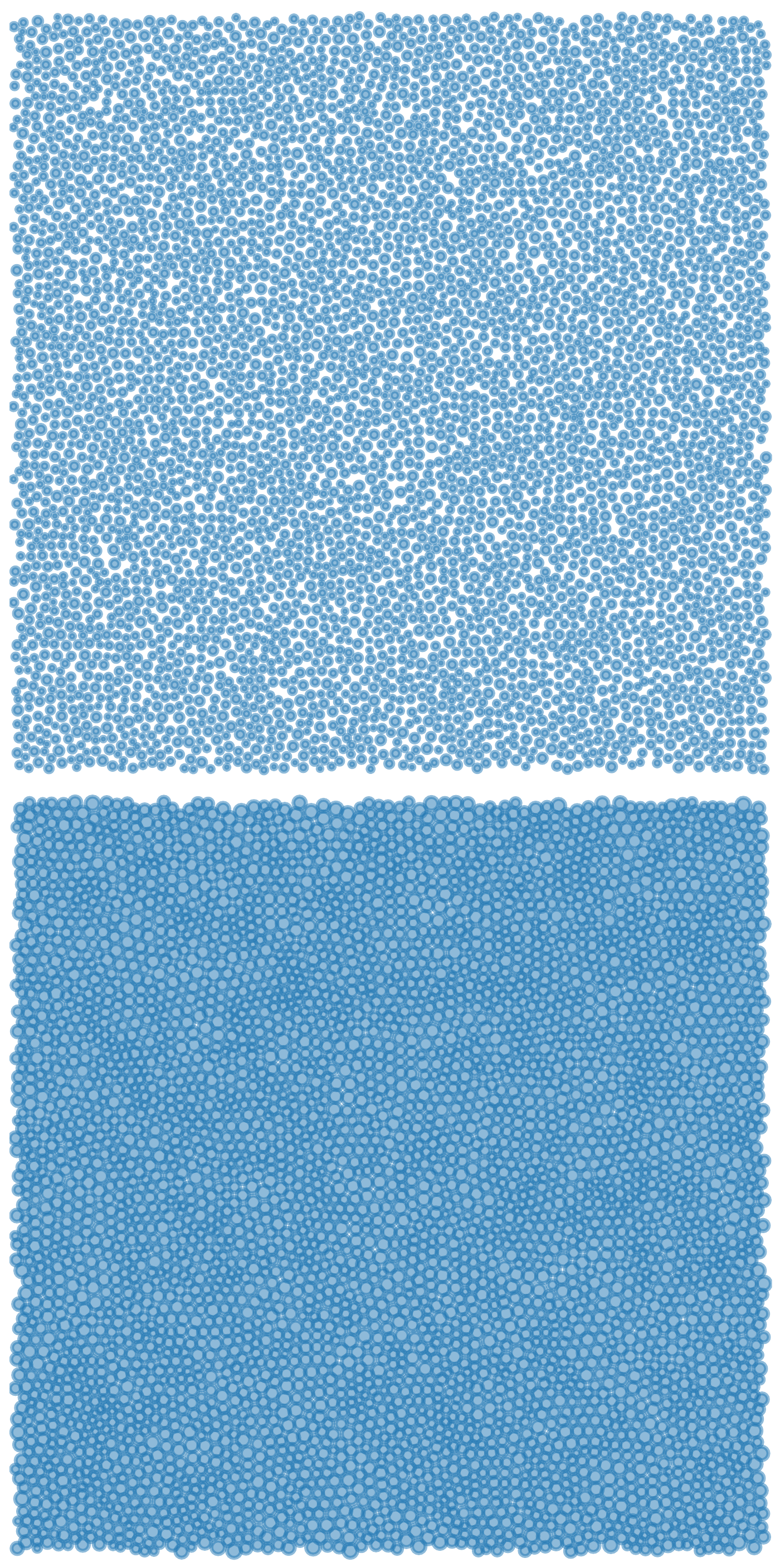}
\caption{\small \textbf{Multi-particle simulations snapshots.} Representative shapshots for the multi-particle simulations for (top) $\phi=0.8$ and (bottom)  $\phi=1.8$ at Hertzian strength $A \approx 680 k_BT$.}
\label{fig:snapsmulti}
\end{figure}

\begin{figure}[t]
\centering
\includegraphics[scale=0.3]{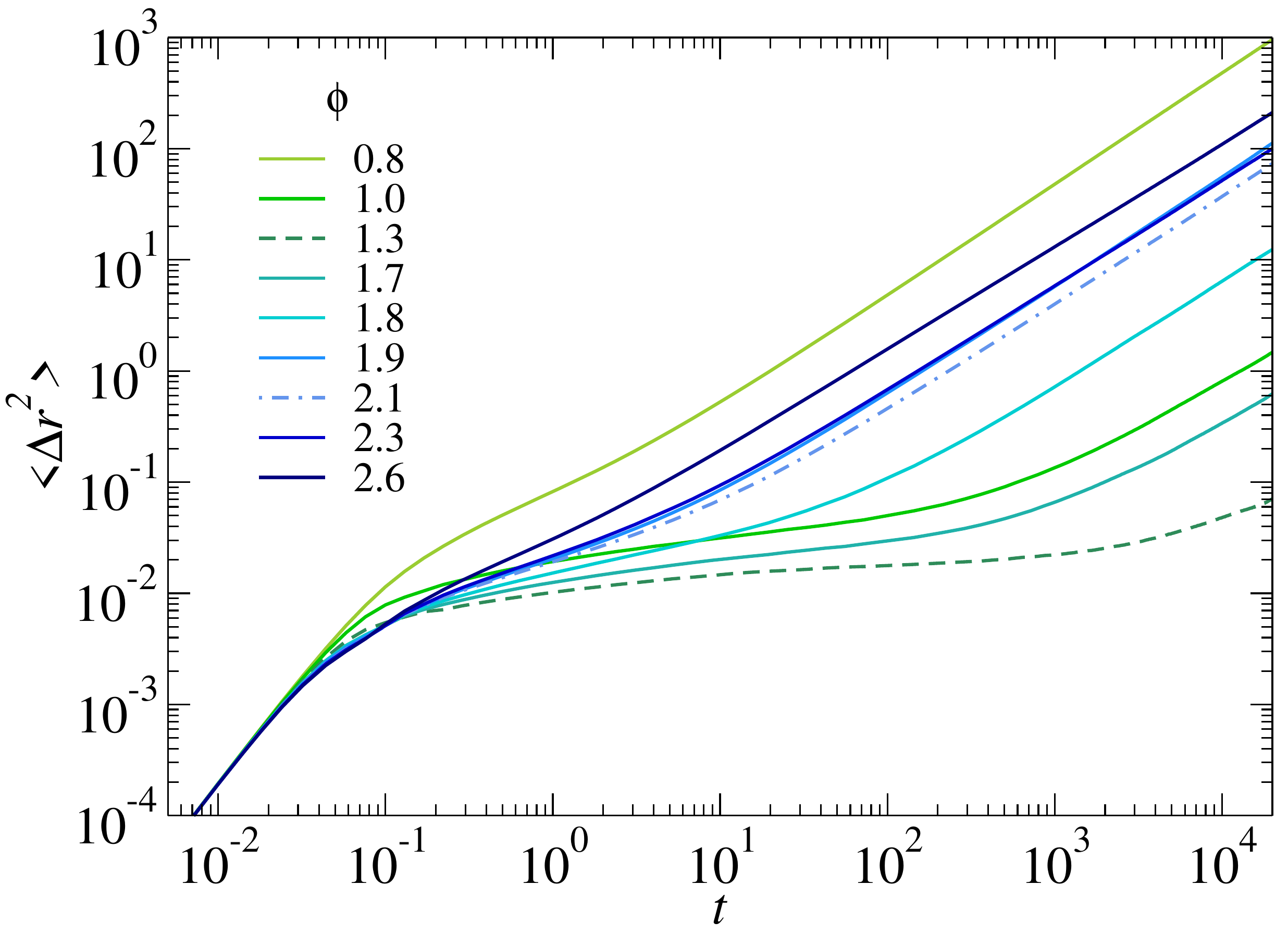}
\caption{\small \textbf{Reentrant behavior.} Mean-squared displacement of disks interacting with the 2D Hertzian potential for different area fractions $\phi$ at a representative value of Hertzian strength $\approx 680 k_BT$. The dashed line signals the onset of a glass while the dashed-and-dotted line highlights a further slowdown of the dynamics at higher $\phi$.}
\label{fig:msd}
\end{figure}

\begin{figure}[H]
\centering
\includegraphics[scale=0.45]{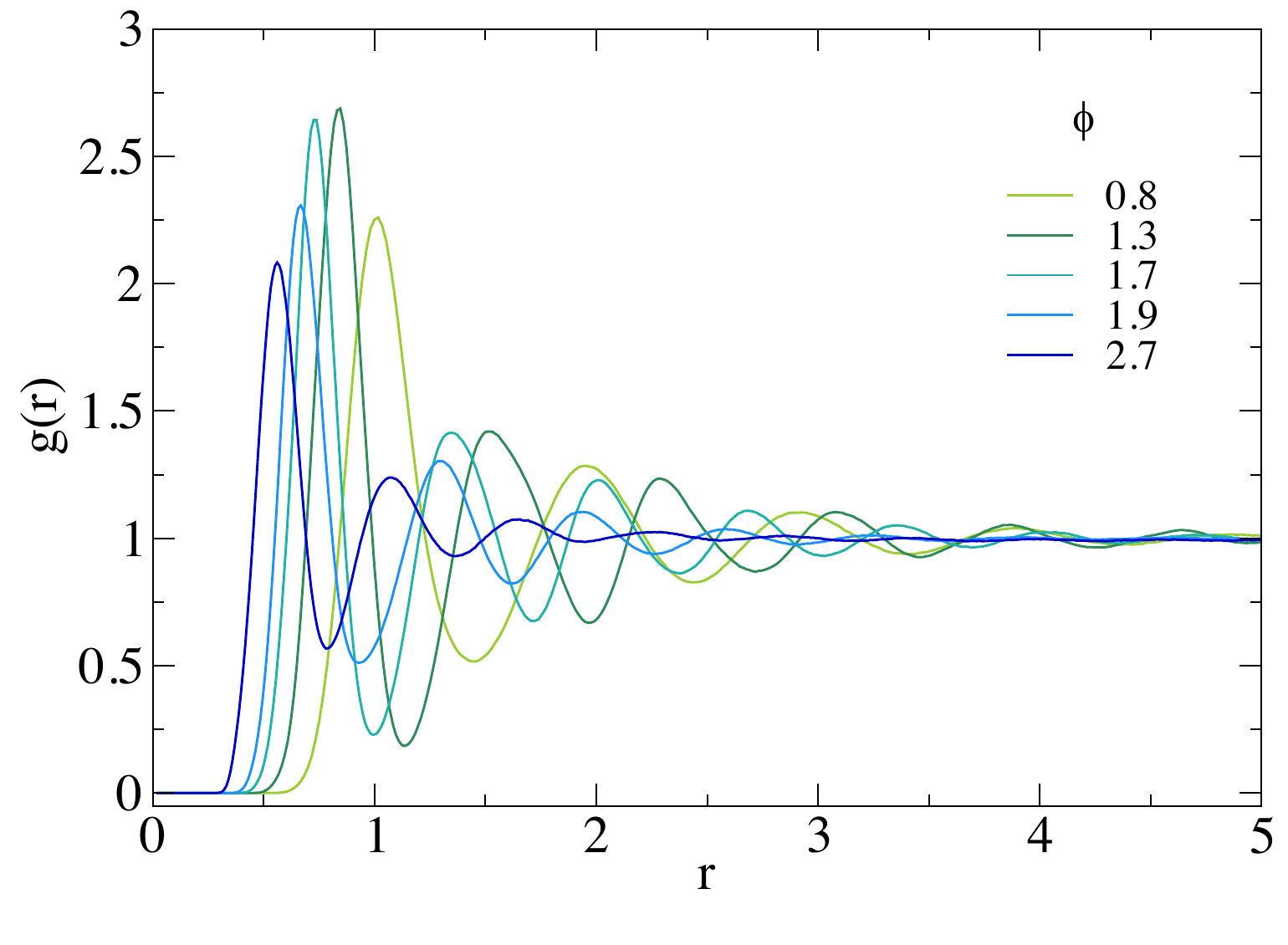}
\caption{\small \textbf{Radial distribution function.} Pair correlation function $g(r)$ as a function of the distance $r$ (measured in units of $\sigma$), for some representative packing fractions  $\phi$ at Hertzian strength $\approx 680 k_BT$ for $p=0.2$.
}
\label{fig:grall}
\end{figure}

Complementing the analysis described in the main text, we report in Fig.~\ref{fig:snapsmulti} two representative simulation snapshots for particles interacting via the 2D Hertian potential with Hertzian strength $\approx 680 k_BT$. The two panels show, respectively, a low $\phi$ system in the fluid regime and a higher $\phi$ one, once the glassy state has been overcome and the ensemble fluidifies again. The snaphots help to visualize the number of overlaps between the particles, that are virtually zero at low packing fraction, but increase rapidly with increasing $\phi$.
\\
\indent
In Fig.~\ref{fig:msd}, we show the mean-squared displacement for the same system for different area fractions. The values of diffusion coefficients $D$ reported in Fig. 4(a) in the main text are extracted by the long-time limit of the mean-squared displacement, as explained in the Methods section. For the present case, $\phi = 1.3$ belongs to the glass region. A non-monotonic behavior of the MSD is observed for area fractions exceeding this value, signaling the presence of multiple reentrant transitions as described in the main text.
\\
\indent
Finally, Fig.~\ref{fig:grall} shows the pair correlation function $g(r)$.
As it is reasonable to expect, by increasing $\phi$ the first peak of the $g(r)$ shifts to lower distances. Its intensity follows the same trend as observed for a three-dimensional system of particles interacting via the 3D Hertzian potential~\cite{berthier2010increasing}.
\\
\\

\bibliography{mybib2}

\end{document}